\documentclass[pra,showpacs,twocolumn,superscriptaddress,floatfix,aps]{revtex4-2}
\usepackage{amsmath,amssymb,bm}
\usepackage{graphicx}
\usepackage{physics}
\usepackage[T1]{fontenc}
\usepackage{comment}
\usepackage{float}
\usepackage{mathtools}
\usepackage[normalem]{ulem}
\usepackage{hyperref}
\usepackage{soul}
\usepackage{braket}
\usepackage[utf8]{inputenc}
\hypersetup{colorlinks,%,%
 linkcolor=blue,%
 citecolor=blue,%
 urlcolor=blue}
\usepackage{color}
\usepackage[dvipsnames]{xcolor}
\usepackage{tikz}
\usetikzlibrary{shapes}

\usepackage{soul,xcolor} %Strike through this text

\newcommand{\sks}[1]{{\color{black} #1}}

\def\dar{\downarrow}
\def\upar{\uparrow}
\begin{document}
\setstcolor{red}

\title{Spin-dependent localization of spin-orbit and Rabi-coupled Bose-Einstein condensates in a random potential} 

\author{Swarup K. Sarkar}
%\email{skanti@iitg.ac.in}
\affiliation{Department of Physics, Indian Institute of Technology Guwahati, Guwahati 781039, Assam, India}

\author{Sh. Mardonov}
\affiliation{New Uzbekistan University, Movarounnahr str. 1, Tashkent 100000, Uzbekistan}
\affiliation{Institute of Engineering Physics, Samarkand State University, University Ave.5, 140104, Samarkand, Uzbekistan}

\author{E. Ya. Sherman}
\affiliation{Department of Physical Chemistry, University of the Basque Country UPV/EHU, 48940 Leioa, Spain}
\affiliation{IKERBASQUE, Basque Foundation for Science, Bilbao, Spain}
\affiliation{EHU Quantum Center, University of the Basque Country UPV/EHU, 48940 Leioa, Spain}

\author{Paulsamy Muruganandam}
%\email{anand@bdu.ac.in}
\affiliation{Department of Physics, Bharathidasan University, Tiruchirappalli 620024, Tamil Nadu, India}

\author{Pankaj K. Mishra}
%\email{pankaj.mishra@iitg.ac.in}
\affiliation{Department of Physics, Indian Institute of Technology Guwahati, Guwahati 781039, Assam, India}
\date{\today}

\begin{abstract}
We investigate the effect of the spin-orbit (SO) and Rabi couplings on the localization of the spin-1/2 condensate trapped in a one-dimensional random potential. Our studies reveal that the spin-dependent couplings create distinct localization regimes, resulting in various relations between localization and spin-related properties. First, we examine the localization in the linear condensate and find that the SO coupling can lead to a transition of the localized state from the ``basin-like'' to the ``void'' region of the potential. For a weak random potential upon an increase in the SO coupling, we find a re-entrant transition from a broad to narrow localized state and back at a higher SO coupling. Further, we analyze the competing role of inter-species and intra-species interactions on the localization of the condensate. We find the appearance of spin-dependent localization as the interactions increase beyond threshold values for a sufficiently strong disorder. Our findings on controlling spin-dependent localization may be useful for future ultracold atomic experiments and corresponding spin-related quantum technologies.

\end{abstract}

\date{\today}

\maketitle

\section{Introduction}
%-------------------------------------
Anderson localization, initially predicted for electrons in weakly disordered media~\cite{Anderson:1958, Melvin:1969, TVR:1979, RevTVR:1985, Ferdinand:2008, PhysRevB.104.214207}, has evolved into a fundamental concept for describing the localization of waves and matter waves across various branches of physics, from condensed matter to quantum optics. This phenomenon has now been realized in a wide class of Physical systems, including optical fields~\cite{Wiersma:1997, Scheffold:1999, Schwartz:2007, Aegerter:07, Topolancik:2007}, microwaves~\cite{Dalichaouch:1991, Dembowski:1999, Chabanov:2000, Pradhan:2000}, and sound propagation~\cite{WEAVER:1990}. Following the successful demonstration of localization in quantum matter waves, such as Bose-Einstein condensates (BECs), in the presence of bichromatic and random potentials in one ~\cite{Billy:2008, Roati:2008}, two ~\cite{White:2020}, and three dimensions~\cite{Skipetrov2008}, BECs have become a nodal point for studying localization in ultracold system. They offer an excellent platform to explore the interplay between interactions, disorder, and quantum coherence in a controlled environment. Subsequent to these experimental breakthroughs, numerous numerical simulations~\cite{Adhikari:2009, Cheng_random:2010, Zhang:2022, Sarkar:2023, Sarkar:2024, Zohra:2024, Kamel:2024},  using mean-field models have been conducted to investigate the complex roles of disorder and interactions in condensate localization.

The interplay between disorder and self-interactions \cite{Kopidakis:2008,Billy:2008,Roati:2008} in the formation of stationary localized states presents a complex challenge due to the involvement of various energy and spatial scales associated with nonlinearities and localization.
The presence of artificial Rabi and spin-orbit (SO) couplings~\cite{Spielman:2011, Campbell:2011} and spin-dependent nonlinearities introduces a new dimension to the rich physics of localization in pseudospin-1/2 condensates. These systems exhibit intriguing competition among fundamental quantum effects, including SO coupling, interactions, and disorder. This interplay holds significant potential for generating novel states of matter and advancing quantum technology applications~\cite{Mossman2019, Valdes2021, Frolian2022}.

The SO coupling significantly affects both the localization of the ground state and its dynamics when confined in the random~\cite{clement:2006} and regular potentials~\cite{Zhang:2023}. For example, using the mean-field Gross-Pitaevskii model, Oztas and Nabiollahi demonstrated that increasing SO coupling generally leads to the delocalization of the condensate in SO-coupled Bose-Einstein condensates (BECs), whereas greater detuning tends to favor localization~\cite{Oztas:2024}. Conversely, Yue \textit{et al.} theoretically showed that SO coupling can substantially reduce backscattering for certain momentum states, which can enhance transport properties~\cite{Yue:2020}.

Similarly to the effect of SO coupling on localization, several numerical studies suggest that increasing repulsive interactions among condensate atoms can induce a transition from localization to delocalization in BECs~\cite{Sanchez:2007, Adhikari:2009, Cheng:2010, Cheng:2011, Sarkar:2023, Sarkar:2024}. In binary BECs, Cheng \textit{et al.}~\cite{Cheng:2010, Cheng:2011} demonstrated that both inter-species and intra-species interactions lead to symmetry breaking, resulting in complex spatial patterns and transitions between different localized states. Similarly, Santos and Cardoso~\cite{Santos:2021} numerically showed that even if one component of a binary BEC is subjected to a quasi-periodic potential, the linear coupling can induce localization in the other component.

The BECs in random potentials exhibit a range of complex phases, including the Bose-glass phase and Lifshitz phase~\cite{Damski2003, Sanchez2007, Sarkar:2023}. These systems also display intricate dynamics, such as spin precession and separation between spin components in the presence of spin-dependent velocities~\cite{Mardonov2015, Mardonov2018, Zhai2023, Xu2024}. The complexity of these localized states poses significant challenges for researchers seeking to develop a comprehensive physical model to understand these phenomena systematically. 

Most studies on localization in random potentials have focused on non-interacting or weakly interacting condensates. However, the intricate interplay between SOC, disorder, and interactions remains unexplored and poorly understood. In this work, we conduct a comprehensive study to analyze how SO coupling, self-interaction, and disorder affect various types of ground state localization in condensates, extending beyond the conventional Anderson localization framework. Our results reveal that SO coupling facilitates transitions in BEC localization between the basin-like and void regions of the random potential. We also demonstrate that increasing inter-species interaction beyond a certain threshold leads to spin-dependent localization-delocalization effects, as the Manakov symmetry is lifted. These findings are in agreement with our analytical models.

%\sks{..whereas, the non-zero value of self-interaction in various components results in spin-dependent delocalization phenomena..}.

The rest of the paper is organized as follows. In Sec.~\ref{sec2} we formulate the model in terms of the Gross-Pitaevskii equations and introduce the main observables and characteristics of the random potential. In Sec.~\ref{withoutinteraction} we consider the effect of SO and Rabi couplings on the localization of the non-self-interacting BEC. In Sec.~\ref{withinteraction} we concentrate on the effect of SO coupling on the spin-dependent localization of self-interacting BEC for two non-Manakov's realizations of nonlinearities. Here we separately consider the effect of the finite inter-species and intra-species interaction on the spin-dependent localization of the condensate.  Finally, in Sec.~\ref{conclusion} we conclude our work.

\section{Mean-field model, observables, and disorder}
\label{sec2}

In this Section, we formulate the main features of the model and discuss their relation to the disorder in one-dimensional systems.

\subsection{Coupled Gross-Pitaevskii equations}
We consider a pseudospin-$1/2$ quasi-1D condensate strongly confined along the transverse direction, which can be modelled using the coupled Gross-Pitaevskii equations (GPEs) as,%
\begin{subequations}%
\label{eqn1}%
\begin{align}%
% \begin{split}
 {\mathrm i} \frac{\partial \psi_{\upar}}{\partial t} = & \left[-\frac{1}{2}\frac{\partial^{2}}{\partial x^{2}} + 
 g_{\upar \upar} \vert \psi_{\upar} \vert ^{2} + 
 g_{\upar\dar} \vert \psi_{\dar} \vert ^{2} + V(x)\right]\psi_{\upar} \notag \\
 & - {\mathrm i} k_{L}\frac{\partial \psi_{\upar}}{\partial x} + \Omega \psi_{\dar}, \label{eqn1(a)}\\ 
 {\mathrm i} \frac{\partial \psi_{\dar}}{\partial t} = & \left[-\frac{1}{2}\frac{\partial^{2}}{\partial x^{2}} + 
 g_{\dar\dar} \vert \psi_{\dar} \vert ^{2} + g_{\dar \upar} \vert \psi_{\upar}\vert^{2} + V(x)\right]\psi_{\dar} \notag \\ &
 + {\mathrm i} k_{L}\frac{\partial \psi_{\dar}}{\partial x} + \Omega \psi_{\upar}, \label{eqn1(b)} 
\end{align}
%\label{eqn1}%
\end{subequations}%
where $\psi_{\upar}$ and $\psi_{\dar}$ $(\psi_{\upar,\dar}\equiv \psi_{\upar,\dar}(x,t))$ represent the pseudo spin-up and spin-down components 
of the condensate wavefunction ${\bm\psi}=\left(\psi_{\upar},\psi_{\dar}\right)^{\rm T}$, respectively, where ${\rm T}$ stands for transposition. For stationary states $\psi_{\upar,\dar}(x,t)=\psi_{\upar,\dar}(x)\exp(-{\mathrm i}\mu t),$ where $\mu$ is the chemical potential common for both spin components. Here $g_{\upar\upar}$ and $g_{\dar\dar}$ are the intra-species nonlinearities for spin-up and spin-down components, respectively, and $g_{\upar\dar}$ represents the inter-species interaction. In what follows, we will study stationary states and omit explicit $x-$dependence for brevity when it will not cause confusion.

The GPEs (\ref{eqn1(a)})-(\ref{eqn1(b)}) correspond to the SO coupling Hamiltonian in the form $-{\mathrm i}k_{L}\sigma_{z}\partial/\partial x$ and the Rabi coupling $\Omega\sigma_{x}$ with $\sigma_{i}\ (i=x,z)$ being the corresponding Pauli matrices. The spin-orbit and Rabi coupling strengths are denoted as $k_{L}$ and $\Omega$, respectively, while the random potential is $V(x).$ \sks{We consider realizations of self-interaction with different inter- and intra-spin nonlinearities, thus, with lifted Manakov's symmetry \cite{Manakov:1974}. The Manakov's symmetry realized at $g_{\upar\upar}=g_{\dar\dar}=g_{\upar\dar}=g_{\dar\upar}$ produces the BEC possessing the spin rotational invariance and considerably simplifies the analysis by gauging away the SO interaction in the absence of the Rabi coupling \cite{Tokatly2010}. This symmetry can be lifted as a result of spin-dependent scattering length in interatomic collisions. As will show in this paper, this lifted symmetry plays a critical role in the spin-dependent BEC localization.}

%\ssg{Need to write about MANAKOV's symmetry and how it can be lifted} \sksr{ We consider realizations with different inter- and intra-spin nonlinearities, where Manakov's symmetry of the spin rotational invariance \cite{Manakov:1974}, assuming that all of them are equal and considerably simplifying the analysis by gauging away the SO coupling in the absence of the Rabi coupling \cite{Tokatly2010}, is lifted.}

To obtain the dimensionless Eq.~(\ref{eqn1(a)})-(\ref{eqn1(b)}), we consider the transverse harmonic oscillator length $a_{\perp} = \sqrt{\hbar/(m\omega_{\perp})}$ as a characteristic length scale with $\omega_{\perp}$ as the transverse harmonic trapping frequency, $\omega_{\perp}^{-1}$ as the timescale and $\hbar \omega_{\perp}$ as the characteristic energy scale. The interaction parameters can be defined in terms of {$g_{{\upar\upar},({\dar\dar})}= 2\mathcal{N}a_{{\upar\upar},({\dar\dar})}/a_{\perp}$,} and $g_{\upar\dar} = 2\mathcal{N}a_{\upar\dar}/a_{\perp}$, where, $a_{{\upar\upar},({\dar\dar})}$ and $a_{\upar\dar}$ represent the intra- and inter-component scattering lengths, respectively, and $\mathcal{N}$ represents the total number of atoms in the condensate. The dimensionless spin-orbit $k_{L} $ and Rabi $\Omega$ coupling are defined in the units of $a_{\perp}^{-1}$ and $2\omega_{\perp}$, respectively. The wavefunction is rescaled with $\sqrt{a_{\perp}}$ and follows the normalization condition $N_{\upar}+N_{\dar}=1$, where
\begin{align}
N_{\upar}=\int_{-\infty}^{\infty} \vert \psi_{\upar} \vert ^{2} dx,\qquad
N_{\dar}=\int_{-\infty}^{\infty} \vert \psi_{\dar} \vert ^{2} dx.
 \label{eqn2}
\end{align}
In addition to the functions $\psi_{\upar}$ and $\psi_{\dar}$, dependent on the SO coupling and self-interaction, it is worth introducing a set of function $\phi_{\alpha}(x)$ (the index $\alpha=0,1,\ldots$, numerates the energy states and $\alpha=0$ corresponds to the ground state in the absence of SO coupling and the self-interactions) describing the spinless states with the eigenenergies $\epsilon_{\alpha}$ in the potential $V(x)$. The corresponding eigenstates in the Rabi coupling have the form 
\begin{align}\label{eq:phix}
{\bm\phi}_{\bar{\alpha}}(x)=\frac{\phi_{\alpha}(x)}{\sqrt{2}}
\left[1,\pm 1\right]^{\rm T} 
\end{align}
with the energies $\epsilon_{\bar{\alpha}}=\epsilon_{\alpha}+\lambda_{\bar{\alpha}}\Omega$, where $\bar{\alpha}=\left(\alpha\vert \lambda_{\bar{\alpha}}\right)$ is a compound index with the second component $\lambda_{\bar{\alpha}}=\pm 1$ corresponding to the $\langle\sigma_{x}\rangle$ value.

\subsection{Definition of observables} \label{sec:IIB}

The localization at different spatial scales is characterized by the spin-projected width $w_{j}$, center of mass $\langle x_{j}\rangle $ and the IPR (inverse participation ratio) $\chi_{j}$ ($j=(\upar,\dar)$) defined as:
\begin{align}\label{eq:wj2}
w_{j}^{2} = \frac{1}{N_{j}}\int_{-\infty}^{\infty} (x-\langle x_{j} \rangle)^{2} \vert \psi_{j} \vert ^{2} dx, 
\end{align}
\begin{align}\label{eq:xm}
\langle x_{j} \rangle = \frac{1}{N_{j}}\int_{-\infty}^{\infty} x \vert \psi_{j} \vert ^{2} dx, \end{align} 
and 
\begin{align}\label{eq:chij}
\chi_{j}=\dfrac{1}{N_{j}^{2}}\int_{-\infty}^{\infty} \vert \psi_{j} \vert ^4 dx,
\end{align}
where $N_j$ is defined by Eq.~(\ref{eqn2}). 

It is instructive for further analysis to present the local density $n_{j}(x)\equiv\left\vert \psi_{j}(x)\right\vert ^{2}$ using the Fourier components $n_{j}(p)$ as:
\begin{align}\label{eq:fourier}
n_{j}(x)=\int_{-\infty}^{\infty} n_{j}(p)\exp({\mathrm i}px) \frac{dp}{\sqrt{2\pi}}
\end{align}
resulting in 
\begin{align}\label{eq:fouroer_chi}
\chi_{j}=\frac{1}{N_{j}^{2}}\int_{-\infty}^{\infty}n_{j}^{2}(p)dp.
\end{align}
Equation \eqref{eq:fouroer_chi} (see also Ref. \cite{Modugno2017}) shows that increasing the contribution of the higher momenta to the probability can increase the IPR. 

We consider spin-dependent observables
\begin{align}\label{eq:spin}
\langle\sigma_{i}\rangle=\int_{-\infty}^{\infty} {\bm\psi}^{\dagger}\sigma_{i}{\bm\psi} dx,
\end{align}
and characterize the purity of the system in the spin subspace as $P=\sum_{i}\langle\sigma_{i}\rangle^{2}$ \cite{Blum:2012}. Here $P=1$ corresponds to a pure state with the spin on the Bloch sphere, while $P=0$ is the fully mixed state with zero length of the spin vector. 

We also utilize the ``spin miscibility'' parameter characterizing the joint distribution of densities of spin components defined as
\begin{align} \label{eq:misc}
\eta=2 \int_{-\infty}^{\infty} \vert {\psi}_{\upar}\vert \vert {\psi}_{\dar} \vert dx,
\end{align}
where $\eta=1$ and $\eta=0$ correspond to fully miscible and immiscible realizations, respectively. For real wavefunctions, we have $\eta=\vert \langle\sigma_{x}\rangle\vert $. Although the purity and spin miscibility are mutually related, they can demonstrate quite different dependence on the system parameters.

\subsection{Characterization of disorder}

To study the evolution of localization of the nonlinear SO coupled condensate in a random potential, we consider the Lifshitz form \cite{lifshits:1988} of the random potential as:
\begin{align}
V(x) =\frac{U_{0}}{\zeta \sqrt{\pi}} \sum_{s = 1}^{{\cal N}_{\rm s}}
\exp (-\frac{(x - x_{\rm s})^{2}}{\zeta^{2}}), \label{eqn:3}
\end{align}
where $U_{0}$ corresponds to the strength of a narrow spike of the width $\zeta$ at uncorrelated random positions $x_{\rm s}$ with ${\cal N}_{\rm s}$ being the total number of spikes. The mean value $\langle V(x)\rangle$ is defined as: 
\begin{align}
\label{eqn7}
\langle V(x)\rangle\equiv \frac{1}{2L}\int_{-L}^L V(x) dx= \bar{n}{U_{0}},
\end{align}
with $\bar{n}={\cal N}_{\rm s}/2L$ being the average concentration of the spikes and $2L$ being the length of the system. 
\sks{To characterize statistical properties of the random potentials in \eqref{eqn:3} common for all the disorder realizations, we introduce the correlation function $C(d)=\langle V(x)V(x+d)\rangle - \langle V(x)\rangle^{2}$ as a function of the distance $d$ between the potential observation points. By assuming an uncorrelated distribution of the positions of the spikes $x_{\rm s},$ we obtain  following Ref. \cite{Shklovskii:1984}:}
\begin{align}
\begin{aligned}
 C(d) &= \langle V(x)V(x+d)\rangle - \langle V(x)\rangle^{2} \\
 &=\left(\langle V^{2}(x)\rangle - \langle V(x)\rangle^{2}\right)
 \exp(-{d^{2}}/{2\zeta^{2}}), %\notag 
\end{aligned}
\end{align}
where $\langle V^{2}(x)\rangle - \langle V(x)\rangle^{2}=\bar{n}U_{0}^{2}/\sqrt{\sqrt{2}\pi}\zeta$. 
%The random distribution of the spikes shows variations with some local concentrations, which can be  considerably different from $\bar{n}$. 

\sks{While $\langle V(x)\rangle$ and $\langle C(d)\rangle$ present $x_{\rm s}-$ realization-independent statistical characteristics of the random potential \eqref{eqn:3}, the random distribution of the spikes can show various realization-dependent patterns of local concentrations, which can localize the BEC beyond the Anderson localization physics.} Broad basin-like domains with 
%$\langle V(x)\rangle_{b}\approx\langle V(x)\rangle$ 
$\left(\langle V(x)\rangle_{b}-\langle V(x)\rangle\right)^{2}\ll\, \bar{n}U_{0}^{2}/\zeta $ (here $\langle V(x)\rangle_{b}$ stands for the potential averaged over the basin region) located between stronger peaks of $V(x)$ produce regions with the effective potential minima which can localize the condensate. Next,let us consider fluctuations with the local concentration of spikes $n<\bar{n}$ of the length $l$ producing local potential $V_{\rm loc}<\langle V(x)\rangle$. For a complete void with $n=0$, the probability of its realization is given by $\exp(-\bar{n}l)$ with the number of voids $\sim L/l\times\exp(-\bar{n} l).$ This void-like fluctuation becomes sufficient to localize the BEC if $l\gtrsim\pi/\sqrt{\langle V(x)\rangle}.$ Since the probability of large basin-like regions or voids is small, for finite length systems, it can strongly depend on the $x_{\rm s}$ realization at all equal other parameters of the disorder. This combination of localization in the basin-like potentials and the voids is critical for the understanding of the spin-related effect in the ground state of the BEC in a random potential. Detailed analysis of localized states in random one-dimensional potentials has been done in Ref. \cite{Azbel:1983}.
$\left(\langle V(x)\rangle_{b}-\langle V(x)\rangle\right)^{2}\ll\, \bar{n}U_{0}^{2}/\zeta $ (here $\langle V(x)\rangle_{b}$ stands for the potential averaged over the basin region) located between stronger peaks of $V(x)$ produce regions with the effective potential minima which can localize the condensate. Next, let us consider fluctuations with the local concentration of spikes $n<\bar{n}$ of the length $l$ producing local potential $V_{\rm loc}<\langle V(x)\rangle$. For a complete void with $n=0$, the probability of its realization is given by $\exp(-\bar{n}l)$ with the number of voids $\sim L/l\times\exp(-\bar{n} l).$ This void-like fluctuation becomes sufficient to localize the BEC if $l\gtrsim\pi/\sqrt{\langle V(x)\rangle}.$ Since the probability of large basin-like regions or voids is small, for finite length systems it can strongly depend on the $x_{s}$ realization at all equal other parameters of the disorder. This localization in the combination of the basin- and void-like domains is critical for the understanding of the spin-related effects in the ground state of the BEC in a random potential. In the absence of the basins and voids, the ground state BEC localization can be governed by the Anderson mechanism. Detailed analysis of localized states in random one-dimensional potentials has been done in Ref. \cite{Azbel:1983}.

The matter wave states in a random potential can be approximately separated into two groups. The first group is well-localized states in the regions of a relatively smaller concentration of the spikes. The ground and low-energy states of our interest belong to this group. The other group is Anderson localized states with the spatial dependence of the form $\phi_{\alpha}(x)\approx\cos(k_{\alpha}x+\varphi_{\alpha})\xi_{\alpha}(x)$, where $\xi_{\alpha}(x)$ is an extended function at the energy-dependent localization length $\ell(k_{\alpha})$ with $k_{\alpha}\ell(k_{\alpha})\gtrsim\,2\pi$, and $\varphi_{\alpha}$ is the corresponding phase. 
For the model of disorder presented by Eq. (\ref{eqn:3}) for slow particles with $k_{\alpha}\zeta\ll\,1$, we obtain the localization length as $\ell(k_{\alpha})\sim k_{\alpha}^{2}/\bar{n}U_{0}^{2}$. As we will show, spin-orbit interaction couples these groups of states and further leads to a sufficient modification in the nature of the localized states.
%%%%%%%%%%%%%%%%%%%%%%%%%%%%%%%%

\begin{figure}[!htp]
\centering
\includegraphics[width=\linewidth]{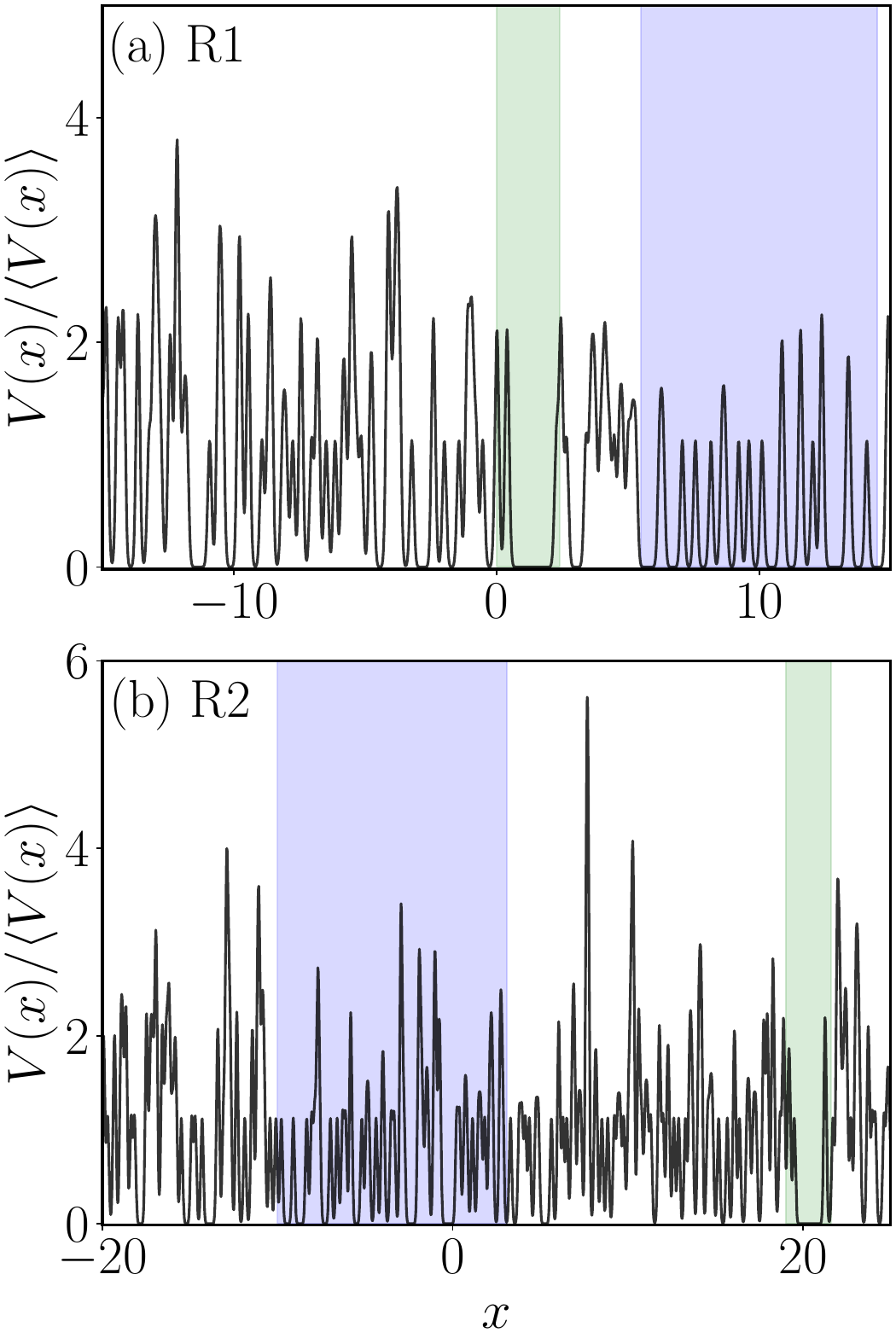}
\caption{The ratio $V(x)/\langle V(x) \rangle$ for realizations (a) R1 and (b) R2, where $\langle V(x)\rangle=\bar{n}U_{0}$ with $\bar{n}=5$ and $\zeta=0.1$. The green and blue shaded regions are drawn respectively to guide the eyes in the void and basin-like regions of the random potential where condensate has a high tendency to get localized. }
\label{fig:potential}
\end{figure}
%%%%%%%%%%%%%%%%%%%%%%%%%%%%%%%%%%%%
\section{Localization of non-interacting BEC}\label{withoutinteraction}

In this Section, we consider BEC without self-interactions, where the GPEs (Eqs.~(\ref{eqn1(a)})-(\ref{eqn1(b)})) become a pair of two coupled linear partial differential equations. In the presence of randomness, the roles of the SO and Rabi couplings become highly nontrivial and can lead to modification of the position and shape of the condensate.

%%%%%%%%%%%%%%%%%%%%%%%%%%%%%%%%%%
 \begin{figure}[!htp]
 \centering
 \includegraphics[width=\linewidth]{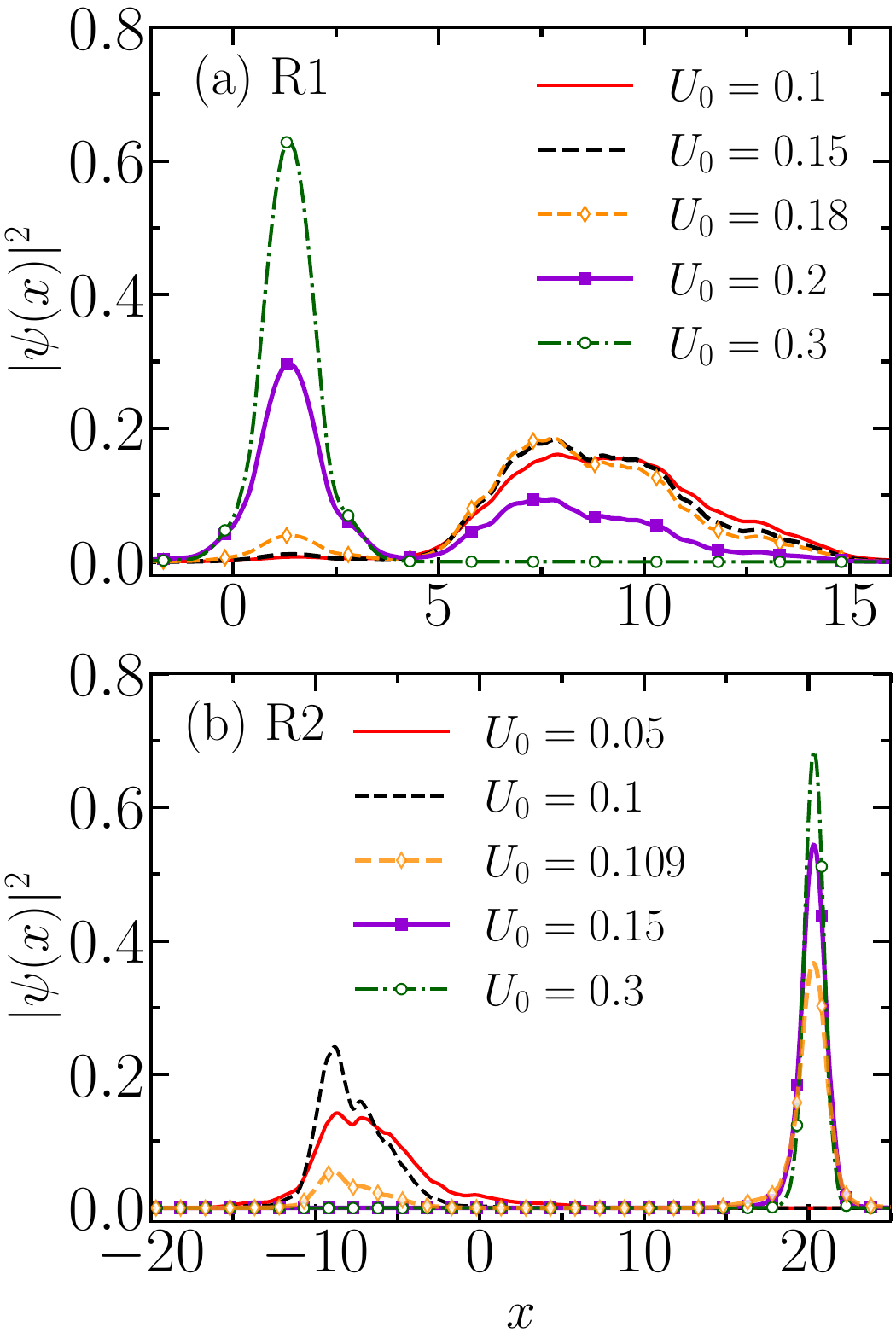}
 \caption{The total density profile for different $U_{0}$ at $k_{L}=\Omega = 0$ for (a) R1 and (b) R2. At small $U_{0}$ the BEC is localized in the basin-like regions of $V(x)$. 
 With the increase in $U_{0}$ it moves to the $V(x)$ voids (cf. Fig. \ref{fig:potential}). 
 Here the interactions are $g_{\upar \upar}= g_{\dar\dar} = g_{\upar\dar} = 0$.}
 \label{fig:density}
\end{figure}
%%%%%%%%%%%%%%%%%%%%%%%%%%%%%%%%%%%%%%%%%%%%%%%%%%%%%%%%%

\subsection{Details of calculation procedure}

We explore the effect of SO and Rabi coupling by analyzing the condensate ground state obtained by solving the coupled linear GPEs \eqref{eqn1} in the potential \eqref{eqn:3} using split-step Crank-Nicholson scheme~\cite{Murug:2009, RAVISANKAR:2021}. We consider the Gaussian wavefunction as the initial state with the imposition of antisymmetric condition given as $\psi_{\upar}(x) = - \psi_{\dar}(-x)$ on between two components. To obtain the localized ground state, we use the imaginary time propagation method aided with split-time Cranck-Nicholson scheme. In the simulation runs presented below, we have considered the spatial steps as $\Delta x = 0.025,$ and time steps as $\Delta t = 10^{-4}.$ To generate the random potential, we choose the number of spikes ${\cal N}_{\rm s} = 512$ within the range $-51.2 < x < 51.2$ with $\bar{n}=5$ and Gaussian width $\zeta = 0.1.$ 

\subsection{Different localized ground state phases in the random potential}

Before analyzing the effect of the SO and Rabi coupling on the BEC localization in detail, we first notice that in the absence of self-interaction, we have $\chi_{\dar}=\chi_{\upar}$, $w_{\upar}=w_{\dar}$, and $\langle x_{\upar}\rangle = \langle x_{\dar}\rangle$. Therefore, in this Section, we use common notations $\chi$, $w$, and $\langle x\rangle,$ respectively, for both components. 

To ensure that the analysis takes into account the realization-dependent properties, wherever relevant we present results for two distinct random realizations labelled as R1 and R2 shown in Fig.~\ref{fig:potential} with the corresponding ground state densities $\phi_{0}^{2}(x)$ shown in Fig.~\ref{fig:density}. 

\begin{figure}[!htp]
 \centering
 \includegraphics[width=\linewidth]{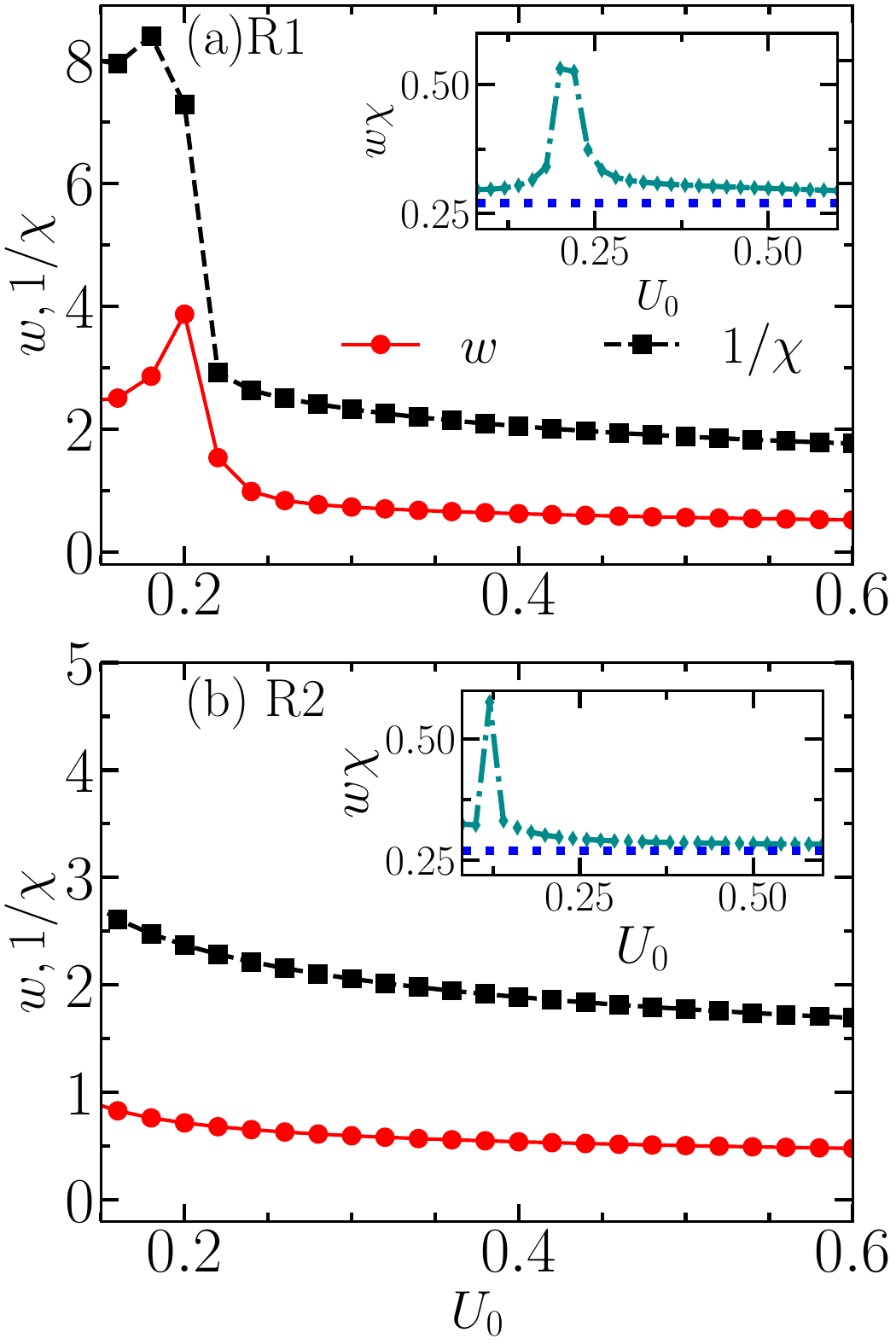}
 \caption{The BEC width $w$ and participation ratio $1/\chi$ as a function of $U_{0}$ for realizations R1 (a) and R2 (b). Insets show the variation of $w\chi$ with $U_{0}$ that asymptotically approaches $w\chi\approx 0.27$ for high $U_{0}$, corresponding to well-defined localization in the voids of the random potential. All the other parameters are the same as in Fig.~\ref{fig:density}. The blue dotted line is drawn at exactly $w\chi = 0.27$. }
 \label{fig:parameters}
\end{figure}

 Comparison of Figs.~\ref{fig:potential} and \ref{fig:density} shows that at
 a small $U_{0}$ the condensate is localized in broad basins of the potential $V(x)$. With the increase in $U_{0},$ a double peak structure including a peak localized in a void of $V(x)$ is being formed by the BEC tunneling between the basin and the void. At a sufficiently large $U_{0}$ the density gets completely confined in this void. Depending on the tunneling probability determined mainly by the basin-void separation, the transition occurs either gradually or sharply with the change in $U_{0},$ at a relatively large and a relatively small tunneling probability, respectively.  These single peak distributions are confirmed by the analysis of the width and the {IPR} presented in Fig.~\ref{fig:parameters}. Upon increasing $U_{0}$, when $\chi$ becomes greater than $\approx 0.6$ and the probability density resembles the localization in a high rectangular potential well. Thus, in the case of R2, the localization towards the void takes place at lower $U_{0}$ compared to R1, but the overall effect of the potential is the same in both realizations.
 
As a result, for the BEC well-localized in a void of the random potential of the width $l$ centered at $\langle x\rangle$, we can use at $\langle x\rangle-l/2<x<\langle x\rangle+l/2$ with the corresponding approximation 
$\phi_{0}^{2}=2\cos^{2}\left(\pi (x-\langle x\rangle)/l\right)/l$, [see Eq. \eqref{eq:phix}] 
and obtain $\chi=3/2l$ and $w=\sqrt{\left(\pi^{2}-6\right)}l/(2\sqrt{3}\pi)$. Thus, the product $w\chi\approx 0.27$, as shown in the insets of Fig.~\ref{fig:parameters} for a relatively large $U_{0}$, while sufficiently larger values 
of $w\chi$ correspond to the double-peak localization at smaller $U_{0}$. 

%%%%%%%%%%%%%%%%%%%%%%%%%%%%%%%%%%%%%%%%%%%%%%%%%%%%%%%
%%%%%%%%%%%%%%%%%%%%%%%%%%%%%%%%%%%%%%%%%%%%%%%%%%%%
%%%%%%%%%%%%%%%%%%%%%%%%%%%%%%%%%%%%%%%%%%%%%%%%%%%%%%%%%%%
\begin{figure}[!htp]
 \centering \includegraphics[width=\linewidth]{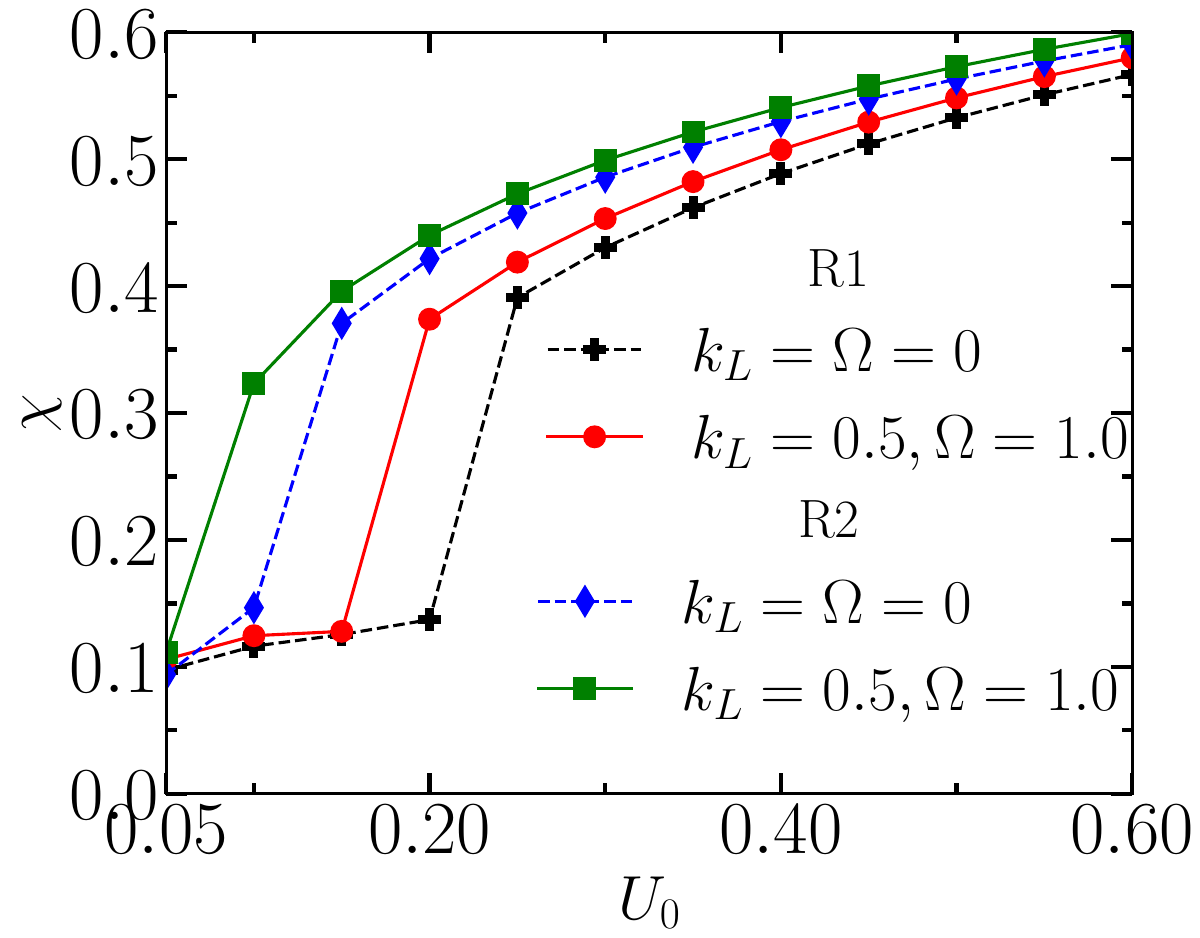}
 \caption{The IPR $\chi$ as a function of $U_{0}$ at $k_{L} = \Omega = 0$ (dashed lines) and $k_{L} = 0.5$, $\Omega = 1.0$ (solid lines) for realizations R1 and R2. The increase in $\chi$ indicates the localization with increasing $U_{0}$ for all cases. Notably, both realizations show slightly higher $\chi$ values in the presence of non-zero SO and Rabi coupling, highlighting the influence of spin-dependent couplings on localization. }
 \label{fig:chi-U0-R1-R2-comp}
\end{figure}
%%%%%%%%%%%%%%%%%%%%%%%%%%%%%%%%%%%%%%%%%%%%%%%%%%%%%%%%%%%%%%%%%%%%%%%%%%%%%%%%
 \begin{figure}[!htp]
 \centering \includegraphics[width=\linewidth]{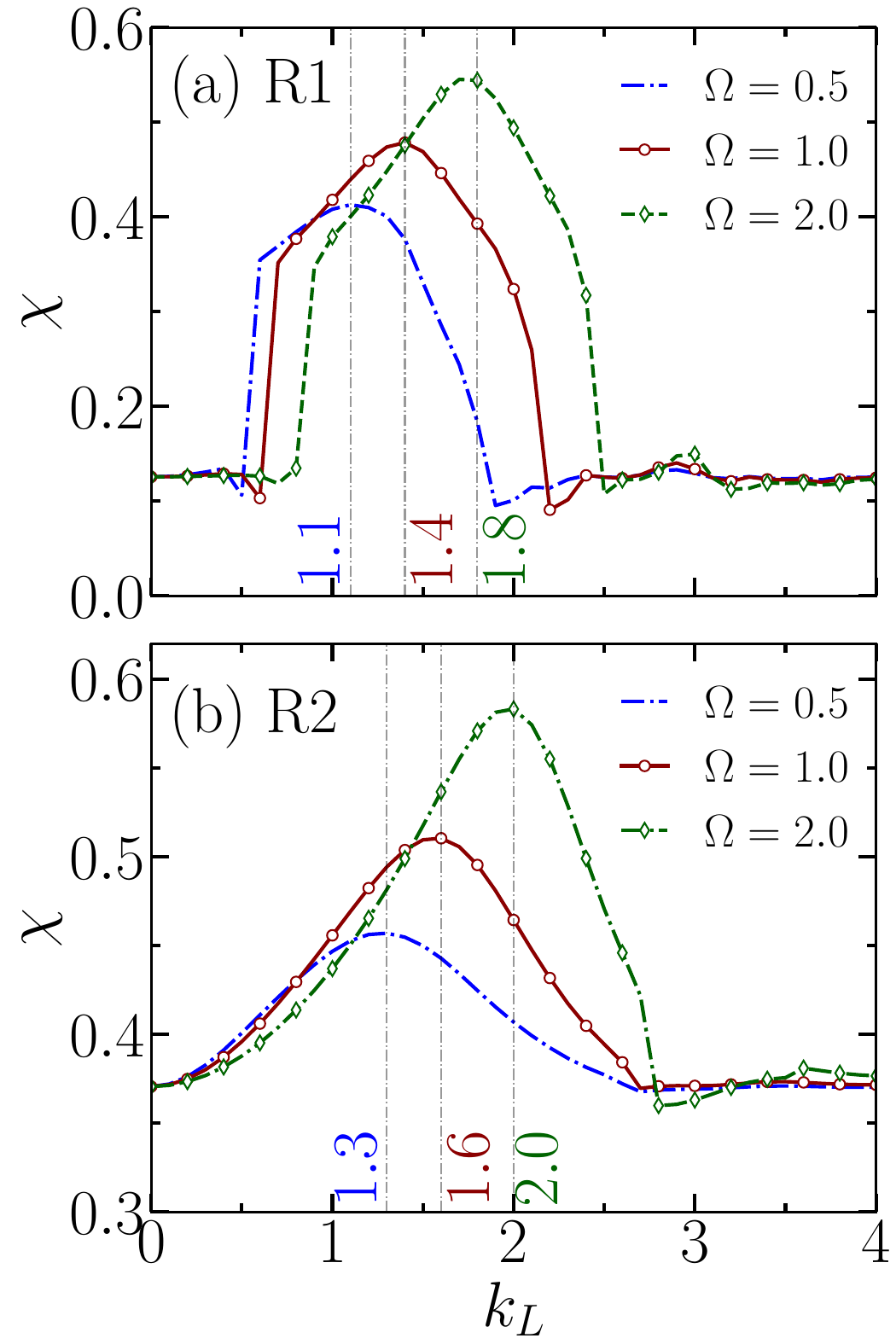}
 \caption{The IPR $\chi$ as a function of SO coupling parameter $k_{L}$ for two realizations as (a) R1, and (b) R2. Here $\Omega = 0.5$ (blue dashed line), $\Omega = 1.0$ (red solid line), and $\Omega = 2.0$ (green dotted line). Upon increasing $k_{L}$, the IPR initially increases and reaches the maximum at intermediate $k_{L}$ (shown by vertical dashed lines) resembling the localization, and then decreases to the $\chi$ value at $k_{L} = 0$. In both cases $U_{0} = 0.15$.  }
 \label{fig:chi-diff-Omega}
\end{figure}
%%%%%%%%%%%%%%%%%%%%%%%%%%%%%%%%%%%%%%%%%%%%%%%%%%%%%%%%%%%%%%%%%%%%%%%%%%%%%%%%
\subsection{Effect of SO and Rabi coupling on the localization for non-interacting BEC}

In order to explore the effect of SO and Rabi couplings, in Fig.~\ref{fig:chi-U0-R1-R2-comp}, we compare the IPR $\chi$ as a function of $U_{0}$ at $k_{L} = 0.5$, $\Omega = 1.0$ with the IPR for $k_{L} = \Omega = 0.0$. Notably, in the presence of $k_{L}$ and $\Omega$ (solid line) the IPR always remains slightly larger compared to the $k_{L} = \Omega = 0$ (dashed line) choice irrespective of the realizations. However, for small $U_{0}$, where $\chi$ is small [Fig.~\ref{fig:parameters}], the differences between the IPRs tend to be significantly larger in comparison to the localized state for higher disorder strength. \sks{Here we notice that in the absence of spin-orbit coupling, Rabi coupling does not influence the BEC localization and {\it vice versa}. In turn, Fig.~\ref{fig:chi-U0-R1-R2-comp} shows that the interplay of the SO and Rabi couplings exerts a more pronounced influence on broad condensate states than on strongly localized ones, as can be analyzed below in this subsection.}

%%%%%%%%%%%%%%%%%%%%%%%%%%%%%%%%%%%%%%%%%%%%%%%%%%%%%%%%%%%%%%%%%%%%%%%%%%%%%%%%%%%%%%%
\begin{figure}[!htp]
 \centering
 \includegraphics[width=\linewidth]{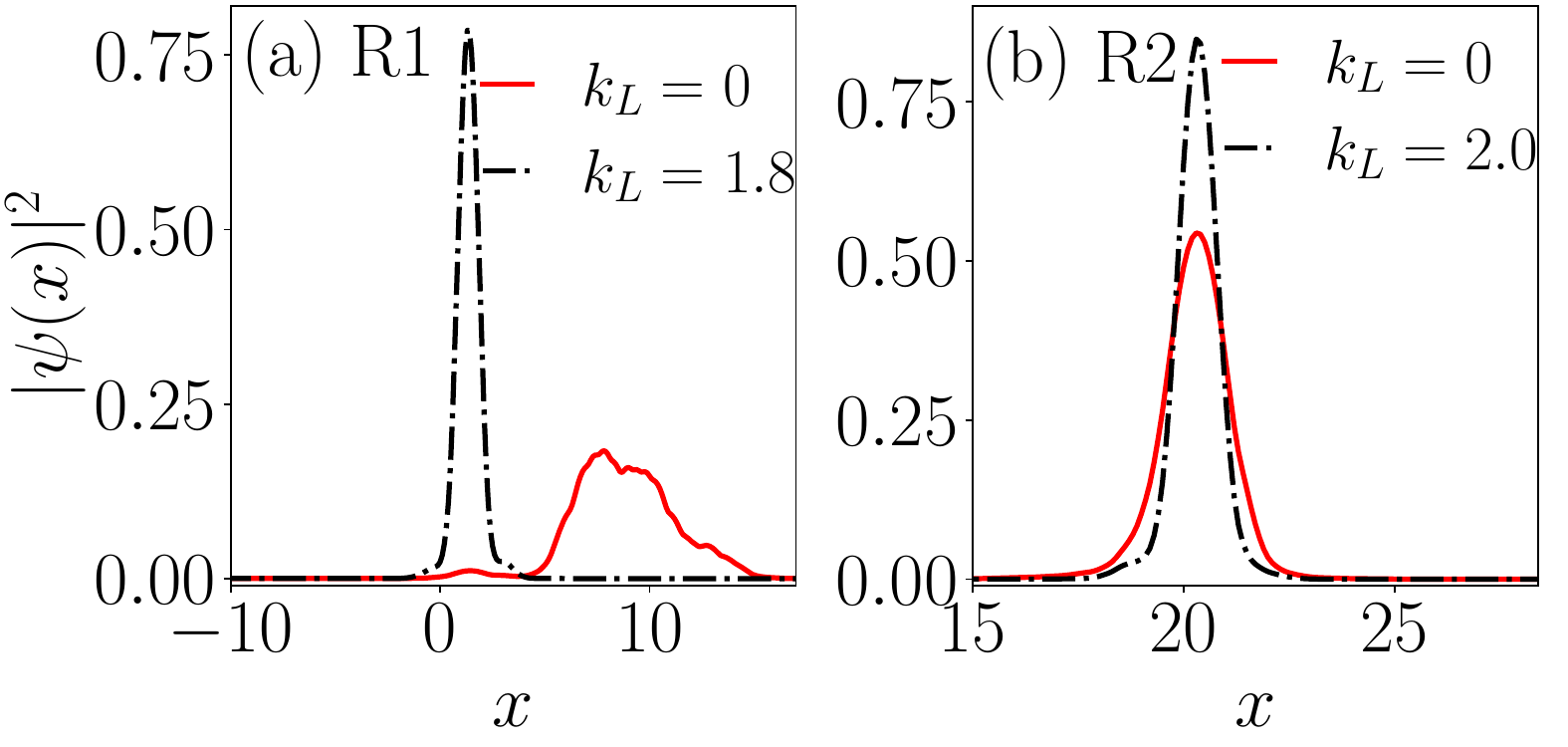}
 \caption{(a) Total density profile for $k_{L} = 0$ (red solid line), and $k_{L} = 1.8$ (black dashed line) which corresponds to the $k_{L}$ at which IPR attains the maximum in Fig.~\ref{fig:chi-diff-Omega}(a) for realization R1. (b) Total density profile at $k_{L} = 0$ (red solid line), and $k_{L} = 2.0$ (black dashed line), the SO coupling associated to the maximum of the IPR in Fig.~\ref{fig:chi-diff-Omega}(b) for realization R2. The other parameters are $\Omega = 2$ and $U_{0} = 0.15$ in both cases.}
 \label{fig:density-kL}
\end{figure}
%%%%%%%%%%%%%%%%%%%%%%%%%%%%%%%%%%%%%%%%%%%%%%%%%%%%%%%%%%%%%%%%%%%%%%%%%%%%%%%%%%%%%%%%
In Fig.~\ref{fig:chi-diff-Omega} we show the variation of $\chi$ as a function of $k_{L}$ at different $\Omega$ for the above R1 and R2 realizations. In Fig.~\ref{fig:chi-diff-Omega}(a) we notice that the IPR remains almost $k_{L}-$independent with $\chi \approx 0.15$ for very low and high value of SO coupling $k_{L} < 0.5$, and $k_{L} > 2.5$, respectively with a steep increase and decrease at certain values of $k_{L}$. Upon increasing Rabi coupling $\Omega$ the amplitude of $\chi$ and the critical value of $k_{L}$ at which the IPR attains maximum shifts to larger values. Similarly, Fig.~\ref{fig:chi-diff-Omega}(b) depicts a similar dependence except that the critical $k_{L}$ values differ from realization R1. In addition to that, in the panel (b) the initial value of $\chi \approx 0.38$ at $U_{0} = 0.15$, corresponding to localization at $k_{L}=0$ in the void (see Fig.~\ref{fig:density-kL}(b)) in comparison to the broadly localized BEC [see Fig.~\ref{fig:density-kL}(a)] for R1. For both the realizations $\chi$ attains maximum at $k_{L}^2 \approx 2 \Omega$.
%%%%%%%%%%%%%%%%%%%%%%%%%%%%%%%%%%%%%%%%%%%%%%%%%%%%%%%%%%
\begin{figure}[!htb]
 \centering
\includegraphics[width=\linewidth]{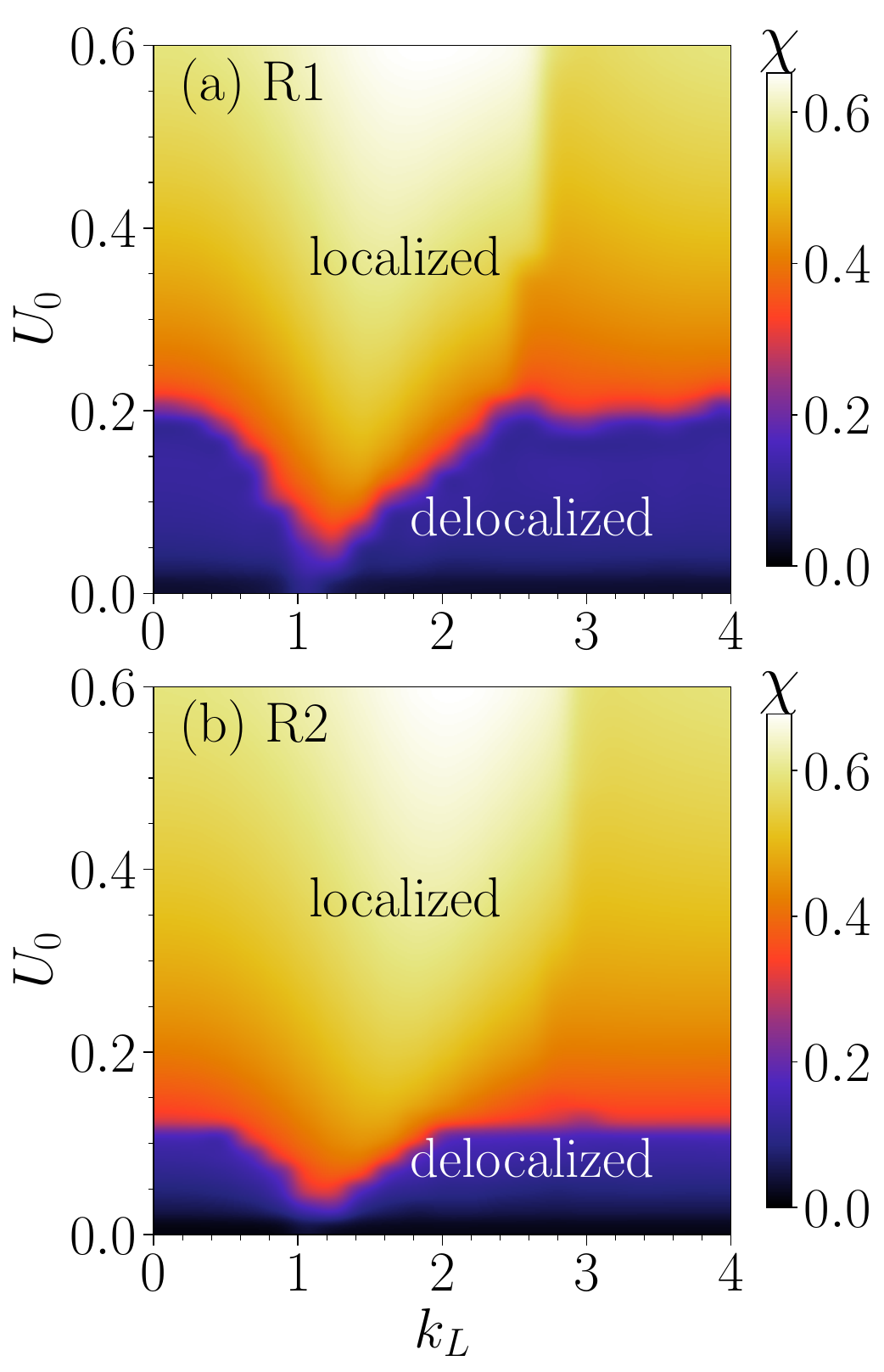}
 \caption{Pseudo-color representation of the IPR $\chi$ in the $(k_{L},U_{0})$ plane keeping $\Omega=1$ for realizations R1 (a) and R2 (b). The localized region is characterized by $\chi \gtrsim 0.3$, while $\chi < 0.3$ indicates the broadly localized condensate, marked as "delocalized" in this and the following Figures. Notably, for both the realizations, with weak disorder strength $U_{0}\approx 0.15$, the condensate gets localized for $k_{L} \approx 1.25$, whereas at higher $k_{L}$ values, it remains broadly localized in the basin. In the case of R2 (b), the transition from broadly to strongly localized state starts at lower $U_{0} \geq 0.1$, compared to $U_{0} \geq 0.2$ for realization R1 in (a). }
 \label{fig:phase-chi-R1-R2-U0-kL}
\end{figure}
%%%%%%%%%%%%%%%%%%%%%%%%%%%%%%%%%%%%%%%%%%%%%%%%%%%%%
Figures \ref{fig:chi-diff-Omega} and \ref{fig:density-kL} present the key results of this Section. The behavior of IPR as a function of $k_{L}$ can qualitatively be understood as follows. For relatively weak SO coupling, one can use perturbation theory \cite{Levitov:2003} with the basis of $\phi_{\bar{\alpha}}(x)$ by calculating modified energies and wavefunctions of the BEC states such that correction to the energy of the state $\bar{\alpha}$ is $\delta\epsilon_{\bar{\alpha}}$. It is important to mention that since the spectrum is dense, after these corrections the order of the index $\alpha$ does not correspond to the order of the energy levels anymore such that inequality $\epsilon_{\bar{\alpha}}+\delta\epsilon_{\bar{\alpha}}>\epsilon_{\bar{\kappa}}+\delta\epsilon_{\bar{\kappa}}$ can be satisfied at $\alpha<\kappa$ and $\lambda_{\bar{\alpha}}=\lambda_{\bar{\kappa}}$.

For the energy correction of the state $\bar{\alpha}$ we obtain by summation over $\bar{\nu}$ eigenstates in the random potential (cf. Eq.\eqref{eq:phix}):
\begin{align}\label{eq:deltaepsilon}
\delta\epsilon_{\bar{\alpha}} = k_{L}^{2}\sum_{\bar{\nu}\neq \bar{\alpha}}
\frac{\vert\,K_{\nu,\alpha}\vert ^{2}}
{\epsilon_{\alpha}-\epsilon_{\nu} +\Omega(\lambda_{\bar{\alpha}}-\lambda_{\bar{\nu}})}, 
\end{align}
where $\lambda_{\bar{\alpha}}\ne\,\lambda_{\bar{\nu}}$ and
\begin{align}\label{eq:alphamu}
K_{\nu,\alpha}=-{\mathrm i}\int_{-\infty}^{\infty} \phi_{\nu}(x)\phi_{\alpha}^{\prime}(x) dx
\end{align}
is the corresponding matrix element of momentum with $\phi_{\alpha}^{\prime}(x)=d\phi_{\alpha}(x)/dx$. 
Similarly, we obtain for wavefunction correction:
\begin{align}\label{eq:deltapsi}
\delta{\bm\phi}_{\bar{\alpha}}= k_{L}\sum_{\bar{\nu}\neq \bar{\alpha}}
\frac{K_{\nu,\alpha}}
{\epsilon_{\alpha}-\epsilon_{\nu} +\Omega(\lambda_{\bar{\alpha}}-\lambda_{\bar{\nu}})}
{\bm\phi}_{\bar{\nu}}.
\end{align}

For a state demonstrating Anderson localization with $\phi_{\nu}(x)\approx\cos(k_{\nu}x+\varphi_{\nu})\xi_{\nu}(x)$ and a localized on the spatial scale $l$ state $\phi_{0}=\sqrt{2}\cos\left(\pi (x-\langle x\rangle)/l\right)/\sqrt{l},$ the momentum matrix element $K_{\nu,\alpha}$ can be presented as:
\begin{align}\label{eq:Kmualpha}
K_{\nu,\alpha}=-\pi\left(\frac{2}{l}\right)^{3/2}k_{\nu}\xi_{\nu}(\langle x\rangle)\frac{\sin(k_{\nu}l/2+\varphi_{\nu})\cos(k_{\nu}l/2)}
{k_{\nu}^{2}-(\pi/l)^{2}}. 
\end{align}
As a result, the Fourier transform of the density acquires higher momentum components resulting (cf. Eq. \eqref{eq:fouroer_chi} and Ref. \cite{Modugno2017}) in the increase in the IPR with the decrease of the spatial scale of the wavefunction. 

The matrix element in Eq. \eqref{eq:Kmualpha} rapidly decreases with the increase in $k_{\nu}$ limiting the summation range over $\nu$ and with the increase in $l$ demonstrating that the perturbation corrections are smaller for broadly localized states. We notice one important observation, specific for the low-energy states in a random potential and, thus, dependent on its realization. The energy correction strongly depends on the shape of the wavefunction and, therefore, on the state number $\alpha.$ Therefore, if perturbed energy $\epsilon_{\bar{\alpha}}+\delta\epsilon_{\bar{\alpha}}$ of a state with $\alpha > 0$  becomes smaller than that of the state with $\alpha=0$ and $\lambda_{\bar{0}}=-1$ (now $\epsilon_{\bar{0}} + \delta\epsilon_{\bar{0}}$), the former becomes the ground state. We can see this in Fig.~\ref{fig:chi-diff-Omega}, where $\chi$ for realization R1 abruptly changes at a certain value of $k_{L}$ and a more localized state becomes the ground state. This is illustrated in the Figures \ref{fig:chi-diff-Omega} and \ref{fig:density-kL}. Figure \ref{fig:chi-diff-Omega}(a) shows a relatively weak effect of SO coupling on the ground state up to a certain value of $k_{L}$ due to its broad localization in a basin of $V(x)$. Then, at a certain value of $k_{L}$ the ground state jumps to a narrow state [see Fig.~\ref{fig:density-kL}(a)] in the $V(x)$ void nearby with this state remaining the ground state in a relatively broad range of $k_{L}$. On the contrary, in the realization R2 [see Fig.~\ref{fig:chi-diff-Omega}(b)] the initial ground state is relatively narrow, the position jump does not occur, and the modification of $\chi$ is smaller than that for the R1 realization. After the jump in the IPR for realization R1, both realizations behave with $k_{L}$ in a similar fashion since the BEC is well localized in a void of $V(x).$

In the strong SO coupling limit one can use linear combinations of degenerate at $\Omega=0$ time-reversed states $\phi_{0}(x)\exp(-{\mathrm i}k_{L}x)[1,0]^{\rm T}$ and $\phi_{0}(x)\exp({\mathrm i}k_{L}x)[0,1]^{\rm T}$ to describe the ground state wavefunction at nonzero $\Omega$. These states have energies $\epsilon_{0} - k_{L}^{2}/2$ and nonzero $\Omega$ produces a doublet of the eigenstates ${\bm\psi}=\phi_{0}(x)\left[\exp(-{\mathrm i}(k_{L}x+\gamma)),\pm\exp({\mathrm i}k_{L}x)\right]/\sqrt{2}$ with the ground state corresponding to $-\exp({\mathrm i}k_{L}x)$ for the spin-down component, where $\gamma$ is the corresponding phase shift. This doublet shows a weak Rabi splitting due to a small spin component 
\begin{align}\label{eq:sigmax}
 \langle\sigma_{x}\rangle = 
 -\left\vert \int_{-\infty}^{\infty}\phi_{0}^{2}(x)\exp(2{\mathrm i}k_{L}x)dx\right\vert .
 \end{align}
 The corresponding $\chi$ parameters here are the same as those at $k_{L}=0$, in agreement with Fig. \ref{fig:chi-diff-Omega}. Due to a small overlap of the spin-up and spin-down components of the ground state spinor, the corresponding spin state is strongly mixed with $P\ll\,1$ while the spin miscibility $\eta$ is still equal to one since their absolute values remain unchanged.

The clear-cut criteria of transition from weak to strong SO coupling for the system cannot be formulated since all involved energies are of the same order of magnitude. A general criterion can be formulated as $k_{L}^{2}\gtrsim\Omega$, similar to the transition from plane wave to stripe phases in disorder-free systems. This condition imposes the upper limit on the transition $k_{L}\sim\Omega^{1/2}$. Another criterion can be formulated as $k_{L}/\chi \gtrsim\,1$, corresponding to a relatively weak spin splitting compared to $2\Omega$, (cf. Eq. \eqref{eq:sigmax}) and this condition is $\Omega-$independent. A complementary scenario is the violation of the lower-order perturbation theory producing relatively weak splitting of the states since the energy shift depends on the spin of the state. States are getting closer in the energy, and this gives a qualitative condition of the transition from weak to strong SO coupling. Since in the system we consider, $\chi$ and $\Omega$ are of the order of 1, all these conditions correspond to the transition at $k_{L}\sim\,1$ also.

Furthermore, to get a comprehensive picture of the parameter-dependent localization scales, we perform an extensive simulation for different ranges of disorder strength $U_{0}$ and SO coupling $k_{L}$ by keeping the Rabi coupling at $\Omega = 1.0$. In Fig.~\ref{fig:phase-chi-R1-R2-U0-kL}, we present the pseudo color plot of the IPR $\chi$ in the $(k_{L}, U_{0})$ plane for realizations R1 and R2. Here, we have used the criteria, $\chi \geq 0.35$ to characterize the condensate in a localized state. In case of R1 [Fig.~\ref{fig:phase-chi-R1-R2-U0-kL}(a)], the IPR remains $\chi \gtrsim 0.4$ and become almost independent of $k_{L}$ after $U_{0} \geq 0.2$, whereas for R2 [Fig.~\ref{fig:phase-chi-R1-R2-U0-kL}(b)], the threshold value of $U_{0}$ above which the $\chi$ is greater than $0.4$, is around $U_{0} \geq 0.1$. However, at $k_{L} \approx 1.25$ the IPR becomes $\chi \approx 0.4$ at a very low value of $U_{0} \approx 0.05$ irrespective of realizations. 

%%%%%%%%%%%%%%%%%%%%%%%%%%%%%%%%%%%%%%%%%%%%%%%%%%%%%
\begin{figure}[!ht]
\centering
\includegraphics[width=\linewidth]{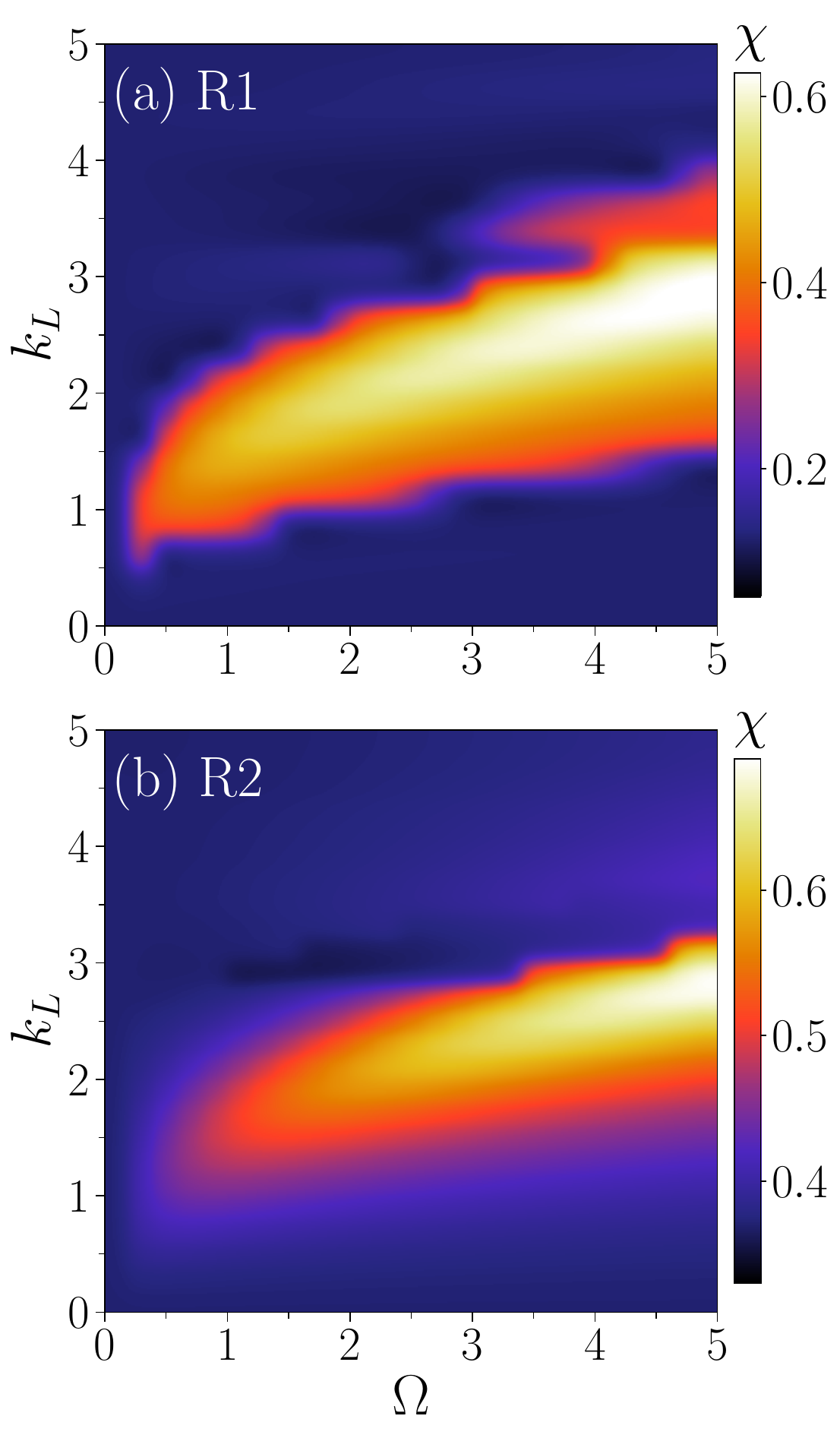}
\caption{Pseudo-color representation of the IPR $\chi$ of the spin-up component in $\Omega - k_{L}$ plane showing the different localization regions for realizations R1 (a) and R2 (b). In panel (a), the broad localization region is characterized with $\chi \leq 0.15$, and the localized region ranges within the range $0.3 \lesssim \chi \lesssim 0.6$. Similarly, the transition boundary in panel (b) is characterized within $0.4 \lesssim \chi \lesssim 0.5$ and in the localized region $\chi \geq 0.5$. }
 \label{fig:phase-kL-Omega-R1-R2}
\end{figure}
%%%%%%%%%%%%%%%%%%%%%%%%%%%%%%%%%%%%%%%%%%%%%%%%%%%%%
%%%%%%%%%%%%%%%%%%%%%%%%%%%%%%%%%%%%%%%%%%%%%%%%%%%%%
\begin{figure}[!ht]
\centering
\includegraphics[width=\linewidth]{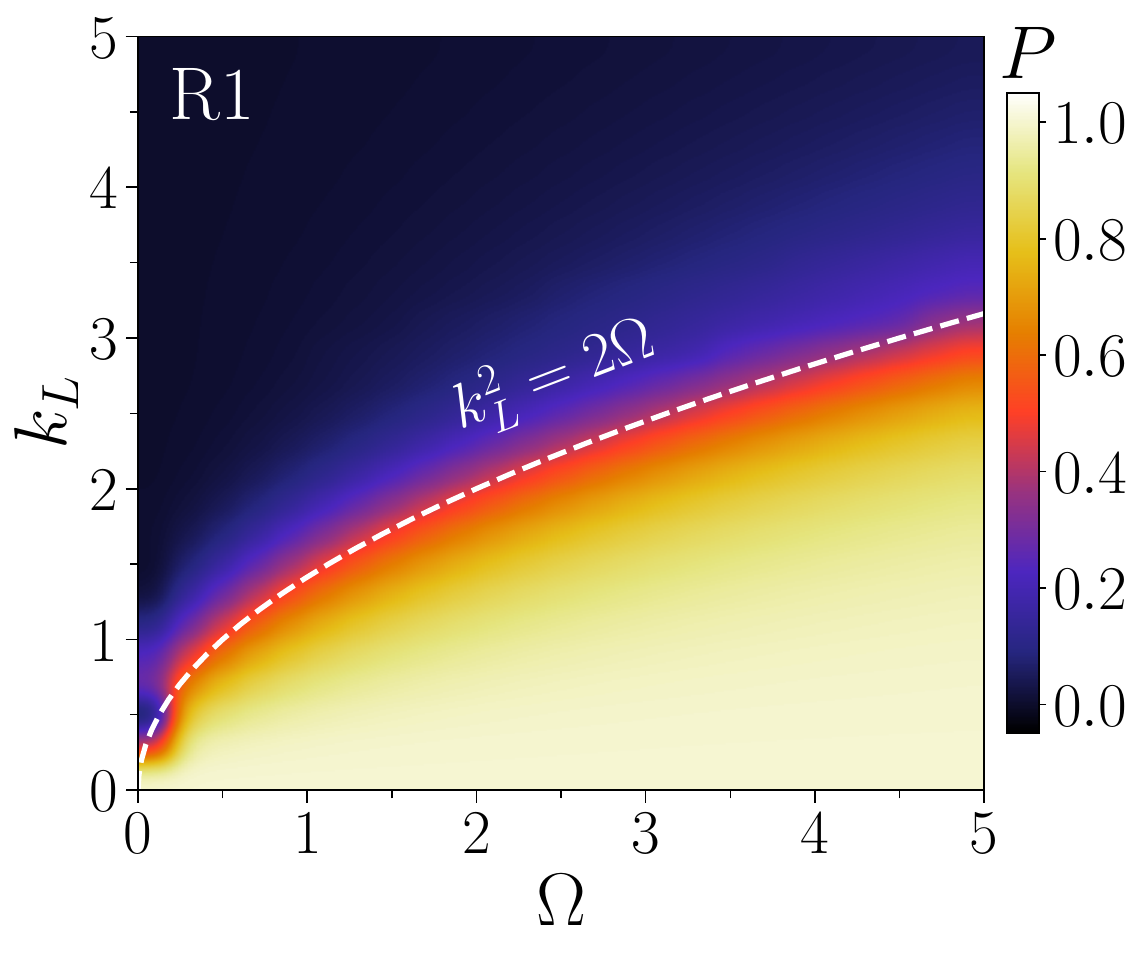}
\caption{Pseudo-colormap representation of purity $P$ in $\Omega-k_{L}$ plane at $U_{0} = 0.15$, $g = 0$ for realization  R1. The white dashed line, drawn at $k_{L}^2 = 2 \Omega$ indicates the transition from a pure to a mixed state, \sks{as the purity $P$ decreases towards 0}.
For R2 realization, this transition follows the same relationship.}
\label{fig:phase-Purity-kl-Omega}
\end{figure}
%%%%%%%%%%%%%%%%%%%%%%%%%%%%%%%%%%%%%%%%%%%%%%%%%%%%%
Next, Fig.~\ref{fig:phase-kL-Omega-R1-R2} illustrates the pseudo color representation of the IPR $\chi$ in $(\Omega,k_{L})$ plane by keeping the disorder strength $U_{0} = 0.15$. In the case of R1 in Fig.~\ref{fig:phase-kL-Omega-R1-R2}(a), the IPR remains at $\chi \lesssim 0.25$ for the broadly localized region at very low and high $k_{L}$ values for a particular $\Omega$. On the other hand, in the localized region, the IPR lies within $0.4 \lesssim \chi \lesssim 0.6$. Using the IPR variation it can be said that the condensate makes a transition from broad to narrow localized and once again it goes to a broadly localized state for higher values of $k_{L} \gtrsim 3.0$ for a fixed $\Omega$. Similar features of the modification in the spatial scale of localization are also observed in Fig.~\ref{fig:phase-kL-Omega-R1-R2}(b) for realization R2, except the amplitude of $\chi$ is larger in comparison to R1 both for narrow- and broadly localized regions. 

Further, we analyze the effect of $k_{L}$ and $\Omega$ on the relation between localization and spin state by calculating the purity $P$ of the condensate. In Fig.~\ref{fig:phase-Purity-kl-Omega}, we show the pseudo color map representation of the purity in $(k_{L}, \Omega)$ plane at $U_{0} = 0.15$ for realization R1. The transition of purity towards zero, maintaining the relation $k_{L}^2 > 2\Omega$ (as marked by a yellow line) signifies the transition from the pure to the mixed phase of the condensate. Note that the condition for this transition is similar to the condition of transition of plane wave to stripe phase of the condensate in disorder-free systems.

So far, we have observed the transition between broad and narrow localized BEC caused by the interplay of the random potential and Rabi, and the SO couplings. To understand the effect of the nonlinearities, in the next Section, we explore the pattern of localization in the presence of intra- and inter-species interactions. 

\section{Localization-delocalization in the presence of self-interaction without Manakov's symmetry}\label{withinteraction}

%\ssg{Need to write about the MANAKOV'S symmetry }\sksr{In this Section, we investigate the effect of the self-interactions on the condensate with lifted Manakov's symmetry (assuming that all inter- and intra-spin interactions are equal), thus, breaking rotational invariance of the BEC self-interaction energy. This absence of the rotational invariance greatly extends the physics of the condensate and will be explored below.}

\sks{In this Section we investigate the effect of the self-interactions without Manakov's symmetry, that is with lifted spin rotational invariance, on the BEC localization. The absence of spin rotational invariance considerably extends the variety of realizations of the condensate localization, as will be shown in the following.}

\subsection{Qualitative analysis}

The Gross-Pitaevskii equations \eqref{eqn1} do not represent an integrable system \cite{Kartashov:2019} making approaches similar to the perturbation theory of the previous Section impossible. However, one can perform a semiquantitative analysis and compare it with the exact numerical results, as it will be done in what follows.

In general, the cross-spin repulsion $g_{\upar\dar}$ and the difference in the self-interaction ($g_{\upar \upar} \neq g_{\dar\dar}$) tend to form different populations $(N_{\upar}\ne N_{\dar})$ by competing with the Rabi coupling. 
To begin with understanding the joint effect of the spin-related couplings and self-interaction we assume that the BEC is near the threshold described by the
trial wavefunction 
\begin{align}\label{eq:psitr}
{\bm\psi}_{\rm tr}=\frac{\phi_{\rm g}(x)}{\sqrt{2}}
\left[\sqrt{1+\varepsilon}\exp({\mathrm i}\theta),-\sqrt{1-\varepsilon}\exp(-{\mathrm i}\theta)\right]^{\rm T}, 
\end{align} 
where $\vert \varepsilon\vert \ll\,1$, and $\phi_{\rm g}(x)$ depend on the random potential, SO coupling, and self-interaction. The position-dependent phase $\theta$ corresponds to the effect of SO coupling.

We expand the total BEC energy in terms of $\varepsilon=N_{\upar}-N_{\dar}$ to the first relevant contributions as 
\begin{eqnarray}\label{eq:Evarepsilon}
\tilde{\epsilon}_{\rm tr}(\varepsilon)&-&\tilde{\epsilon}_{\rm tr}(0) = \frac{\chi}{4}
\left(g_{\upar\upar}-g_{\dar\dar}\right)\varepsilon + \\
&&\left[
\frac{\chi}{8}\left(g_{\upar\upar}
-g_{\dar\dar}\right)-\frac{\Omega}{2}\langle \sigma_{x}\rangle - 
\frac{\chi}{4} g_{\upar\dar}
\right]\varepsilon^{2}, \notag
%&&+ \frac{k_{L}}{\sqrt{2}}\varepsilon
%\int_{-\infty}^{\infty}\phi_{\rm g}^{2}%(x)\left(2\theta_{\upar}\right)^{\prime} dx,
 %\notag
\end{eqnarray}
where
\begin{align}\label{eq:sigmaxtheta}
\langle \sigma_{x}\rangle = -\int_{-\infty}^{\infty}\phi_{\rm g}^{2}(x)\cos\left(2\theta\right) dx,
\end{align}
and the interaction-dependent $\chi$ is the same for both spin components. In Eq.\eqref{eq:Evarepsilon}
\begin{align}\label{eq:Evarepsilon0}
\tilde{\epsilon}_{\rm tr}(0)=\tilde{\epsilon}_{\rm lin} + 
\frac{\chi}{8}\left(g_{\dar\dar}+g_{\upar\upar}\right)+
\frac{\chi}{4}g_{\upar\dar},
\end{align}
where $\tilde{\epsilon}_{\rm lin}$ is the contribution of the terms not including self-interactions explicitly.

One can draw several following conclusions from Eqs. \eqref{eq:Evarepsilon}, \eqref{eq:sigmaxtheta}, and \eqref{eq:Evarepsilon0}. 

(i) At $k_{L}=0$ and $g_{\dar\dar}=g_{\upar\upar}=0$ to have the effect on the BEC, the repulsion $g_{\upar\dar}$ should exceed a certain threshold with $g_{\upar\dar}\chi>2\Omega$. Thus, a competition between $\Omega$ and $g_{\upar\dar}\chi$, makes the effect of cross-spin repulsion relatively weaker for extended states with $\chi\ll\,1$. At $g_{\dar\dar} = g_{\upar\upar}$ the energy depends on $\varepsilon^{2}$, making the ground state double degenerate while any $g_{\dar\dar}\ne\,g_{\upar\upar}$ lifts this degeneracy.

(ii) At $k_{L}=0$ and nonzero $g_{\dar\dar}-g_{\upar\upar}$, the transition for spin inequality has no threshold since same spin repulsion yields a term linear in $\varepsilon$ in the energy and the resulting  $N_{\upar}-N_{\dar}$ depends on $g_{\dar\dar}-g_{\upar\upar}.$. 

(iii) Self-repulsion expands the BEC when the self-interaction energy becomes close to the ground state energy in the non-self-interacting system (e.g., Ref. \cite{Li:2024}). Thus, nonlinearities decrease $\chi_{\upar}$ and $\chi_{\dar}$, making them different, as it depends on the self-interaction parameters and causing spin-dependent broadening. Different density profiles for spin-up and spin-down components contribute to all their spin-related properties.

(iv) Spin-orbit coupling in one-dimensional systems has a complex role in the presence of nonlinearities and disorder \cite{Kartashov:2019}. Firstly, it modifies the mean value of $\langle\sigma_{x}\rangle$ by adding phases to the wavefunction components. Secondly, it can increase the role of the self-interaction producing $z-$axis spin polarization since fully $z-$polarized spin states acquire the contribution of $-k_{L}^{2}/2$ to the total energy. Thus, one can expect its critical role in the presence of nonlinearities, as we will explore further below.

To investigate numerically the effect of spin-related interaction and the nonlinearities on the localization of the condensate, we explore below two scenarios. Firstly, we tune the inter-species repulsion $g_{\upar\dar}$, and secondly, vary the intra-species interaction $g_{\upar\upar(\dar\dar)}$. In both cases, we show that the inequality $g_{\upar\upar} \neq g_{\dar\dar}$ is critically important for the BEC properties.
%%%%%%%%%%%%%%%%%%%%%%%%%%%%%%%%%%%%%%%%%%%%%%%%%%%
\begin{figure}[!ht]
 \centering
 \includegraphics[width=\linewidth]{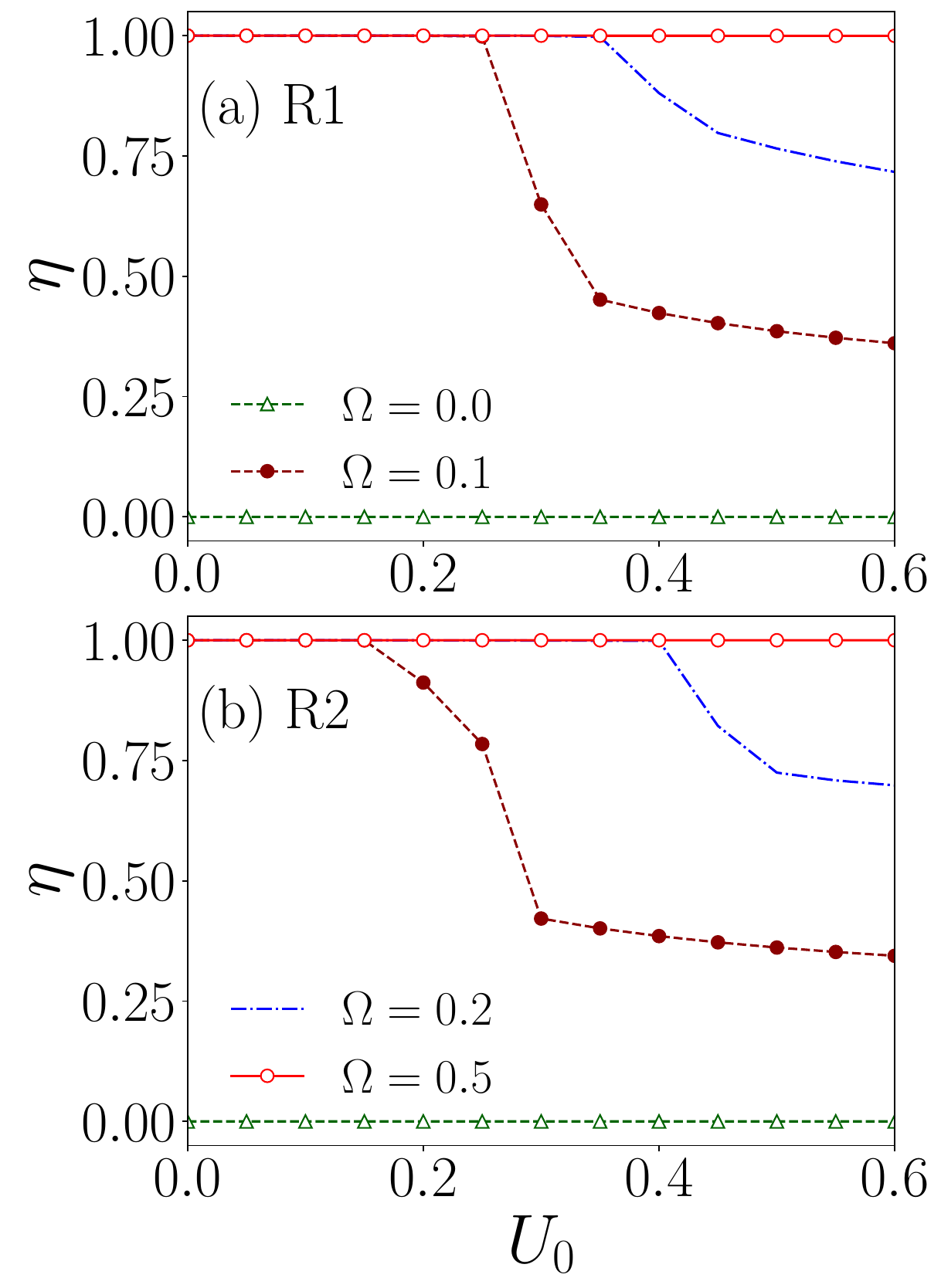}
 \caption{Variation of miscibility $\eta$ with disorder strength $U_{0}$ for different Rabi-coupling $\Omega$ in case of realizations R1 (a), and R2 (b). For $\Omega \leq 0.2$, $\eta$ decreases with the increment of $U_{0}$ beyond a threshold value of $U_{0}$. In the case of R2, the threshold value for $\eta \ll 1$ is lower than R1. Here, other parameters are $g_{\upar \upar} = 0.1$, $g_{\dar\dar} = 0.5g_{\upar \upar}$, $g_{\upar\dar} = 1.0$, and SO coupling $k_{L} = 0$.}
 \label{fig:eta-U0-diff-Omega}
\end{figure}
%%%%%%%%%%%%%%%%%%%%%%%%%%%%%%%%%%%%%%%%%%%%%%%%%
%-------imiscible phases----------
\subsection{Effect of the Rabi coupling and disorder strength on the spin miscibility}

%%%%%%%%%%%%%%%%%%%%%%%%%%%%%%%%%%%%%%%%%%%%%%%%%%%%%
\begin{figure}[!htp]
\centering
\includegraphics[width=\linewidth]{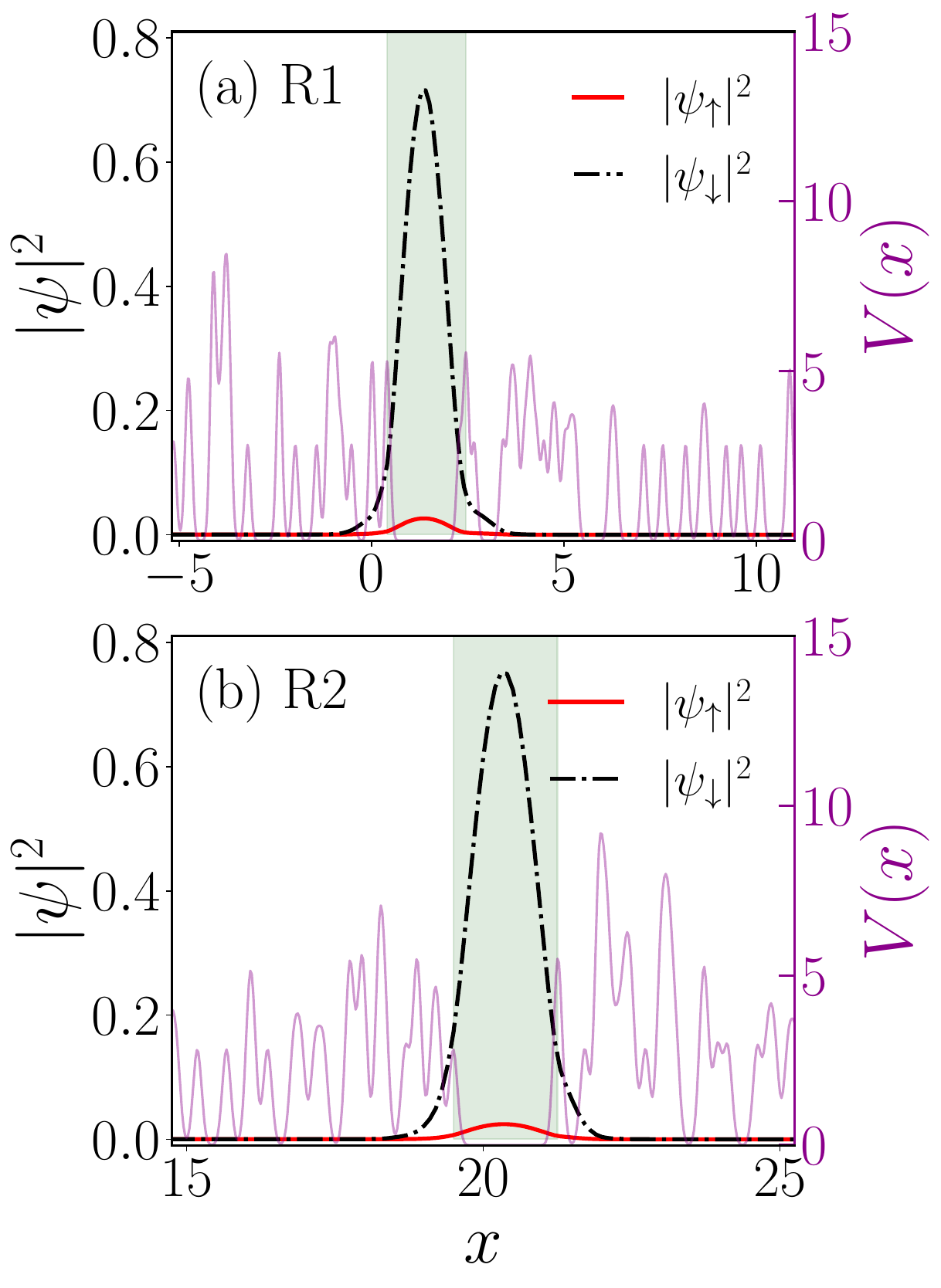}
\caption{Condensate density at $\Omega = 0.1$, $U_{0} = 0.5$. The interaction and SO coupling parameters are $g_{\upar \upar} = 0.1$, $g_{\dar\dar} = 0.05$, $g_{\upar\dar} = 1.0$, and $k_{L} = 0$. The random potential profile is shown with light violet lines along the right side axis. The shaded region is drawn to show the void of the potential.}
\label{fig:density-Omega-0p1-R1-R2}
\end{figure}
%%%%%%%%%%%%%%%%%%%%%%%%%%%%%%%%%%%%%%%%%%%%

After introducing at $k_{L}=0$ a small imbalance between $g_{\upar\upar}$ and $g_{\dar\dar}$ to break the symmetry of the Hamiltonian, the competition between the inter-species interaction $g_{\upar\dar}$ and the Rabi coupling $\Omega$ significantly affects the spin miscibility $\eta$ (see Eq.~\eqref{eq:Evarepsilon}). In addition to that, the random potential plays a critical role in this parameter.

To investigate the effect of the disorder, in Fig.~\ref{fig:eta-U0-diff-Omega} we show the $\eta$ [see Eq.~\ref{eq:misc}] as a function of $U_{0}$ for different $\Omega$ for the realizations R1 and R2. For both realizations, the miscibility $\eta \equiv 0$ for $\Omega = 0$ since here we have $\psi_{\upar}\equiv\,0$. At $\Omega>0$ and small $U_{0}$ one still obtains $\eta=1$ since the product $g_{\upar\dar}\chi$ 
for extended states is small and cannot overcome the effect of the Rabi splitting. After a certain threshold in $U_{0}$, corresponding to the condition $g_{\upar\dar}\chi>2\Omega$, the spin miscibility decreases as expected from Eq. \eqref{eq:Evarepsilon} and \eqref{eq:sigmaxtheta}. This threshold value of $U_{0}$ increases with increasing $\Omega$ up to a certain range, beyond that $\Omega$ the condensate remains perfectly miscible for higher Rabi coupling. The threshold values $U_{0}$ in the case of R2 (Fig. \ref{fig:eta-U0-diff-Omega} panel (b)) are lower compared to R1, which complements the analysis of Fig.~\ref{fig:phase-chi-R1-R2-U0-kL}. After reaching the spin population imbalance, the inter-species repulsion can spatially separate spin components leading to $\langle x_{\upar}\rangle \ne \langle x_{\dar}\rangle.$ Following the decay, $\eta$ achieves a plateau-like behavior with a small slope, where it remains approximately constant while the plateau amplitude decreases with the increase in $\Omega.$ Conclusively, the influence of the Rabi coupling towards spin mixing can effectively be reduced with the increase of the disorder strength as illustrated in Fig.~\ref{fig:density-Omega-0p1-R1-R2}. We notice that the redistribution of population from the spin-up $\vert \psi_{\uparrow}\vert ^2$ to the spin-down $\vert \psi_{\downarrow}\vert ^2$ takes place, in which both components still remain in the void of the potential due to a relatively weak inter-species repulsion. This phenomenon can be attributed to the {\it induced localization} as recently has been observed in~\cite{Santos:2021}.

A qualitative description of two features of this plateau-like behavior related to the fact that it is characterized by the $N_{\upar}\ll\, N_{\dar}\approx\,1$ inequality is in the order here. First, $\psi_{\dar}\approx\,\phi_{0}$ in the given potential $V(x)$ such that the wavefunction can be presented as ${\bm\psi}=[\sqrt{N_{\upar}}\psi_{\upar},\sqrt{N_{\dar}}\phi_{0}]^{\rm T}$, where $\psi_{\upar}$ satisfies the GPE \eqref{eqn1(a)} 
with $\psi_{\dar}\approx\phi_{0}$ (cf. Eq. \eqref{eq:phix}). Next, taking into account that $\chi_{\dar}\sim\chi_{\upar}\sim\chi$ we find that the spin miscibility in Eq. \eqref{eq:misc} shows $\eta\sim\sqrt{N_{\upar}}\gg\,N_{\upar}$. Minimization of the total energy with spin-dependent contributions yields $N_{\upar}\sim (\Omega/(g_{\upar\dar}\chi))^{2}$, and therefore $\eta\sim\Omega/(g_{\upar\dar}\chi)\gg N_{\upar}$.

\subsection{Effect of finite inter-species interaction $g_{\upar\dar}$ on the localization}

%%%%%%%%%%%%%%%%%%%%%%%%%%%%%%%%%%%%%%%%%%%%%%%%%%%%
\begin{figure}[!ht]
\centering
\includegraphics[width=\linewidth]{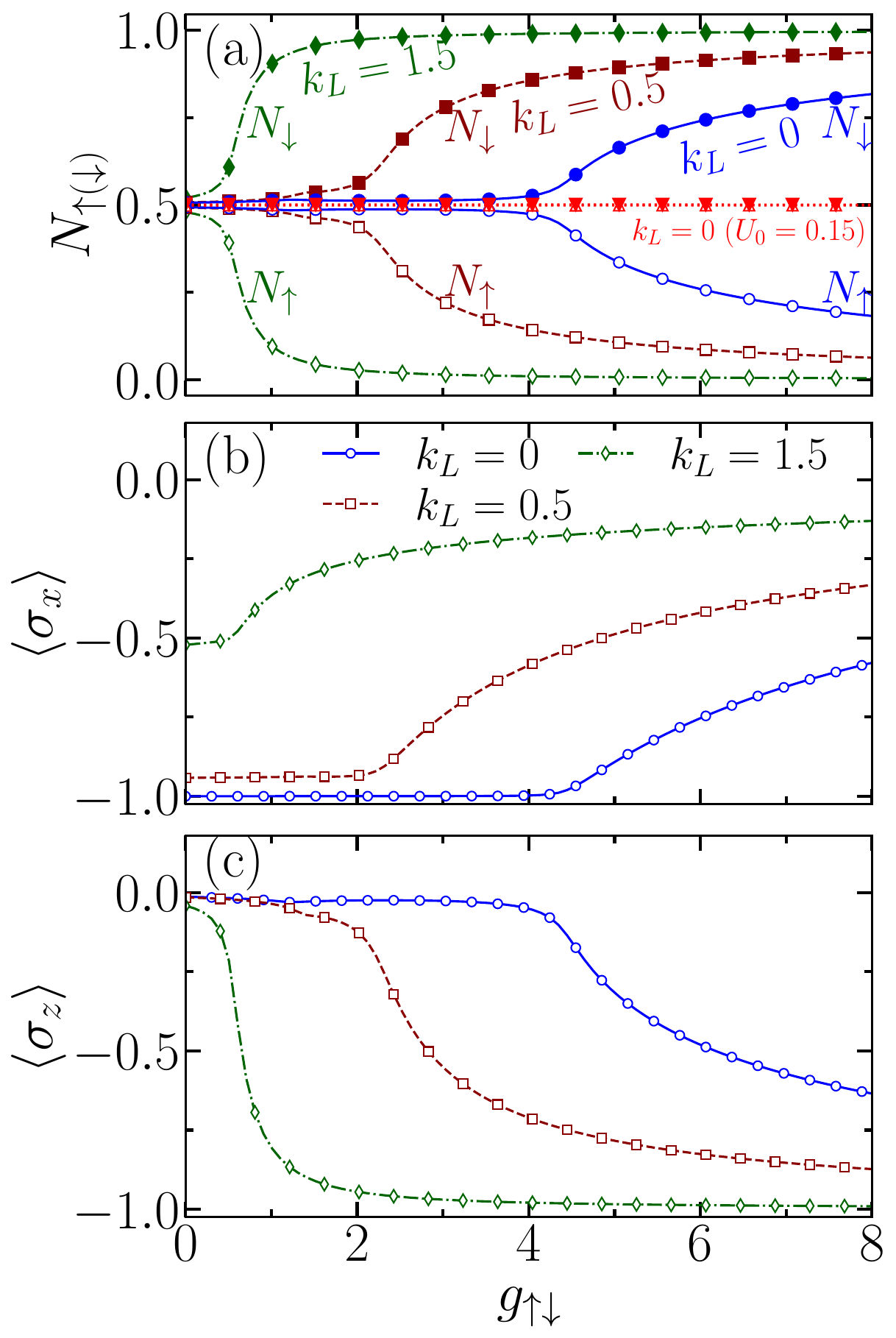}
\caption{(a) Variation of populations with inter-species interaction $g_{\upar\dar}$ by keeping intra-species interactions at $g_{\upar \upar} = 0.1, g_{\dar\dar} = 0.5 g_{\upar \upar}$, and the coupling parameters are \sks{$k_{L} = 0$ (blue circle marked solid line), $k_{L} = 0.5$ (brown square marked dashed line), and $k_{L} = 1.5$ (green diamond marked dash-dotted line)} at $\Omega = 0.5$. Since $g_{\upar \upar}>g_{\dar\dar}$, owing to repulsion, we obtain $N_{\upar}<N_{\dar}$. \sks{Here, we also show the populations for $U_{0} = 0.15$, $k_{L} = 0$ (red triangles with dotted line) in order to compare it with $N_{\upar(\dar)}$  at $U_{0} = 0.5$ (blue solid line).} Variations of $\langle \sigma_{x} \rangle$  and $\langle \sigma_{z} \rangle$ as a function of inter-species interaction $g_{\upar\dar}$ for $k_{L} = 0$ (blue solid line), $0.5$ (brown dashed line), and  $1.5$ (green dash-dotted line) at $U_{0} = 0.5$ are shown in panels (b) and (c), respectively.}
 \label{fig:Natoms-g12}
\end{figure}
%%%%%%%%%%%%%%%%%%%%%%%%%%%%%%%%%%%%%%%%%%%%%%%%%%%%%%%%%%%%%%%%%%%
%%%%%%%%%%%%%%%%%%%%%%%%%%%%%%%%%%%%%%%%%%%%%%%%%%%%%%%%%%%%%
\begin{figure}[!ht]
 \centering
\includegraphics[width=\linewidth]{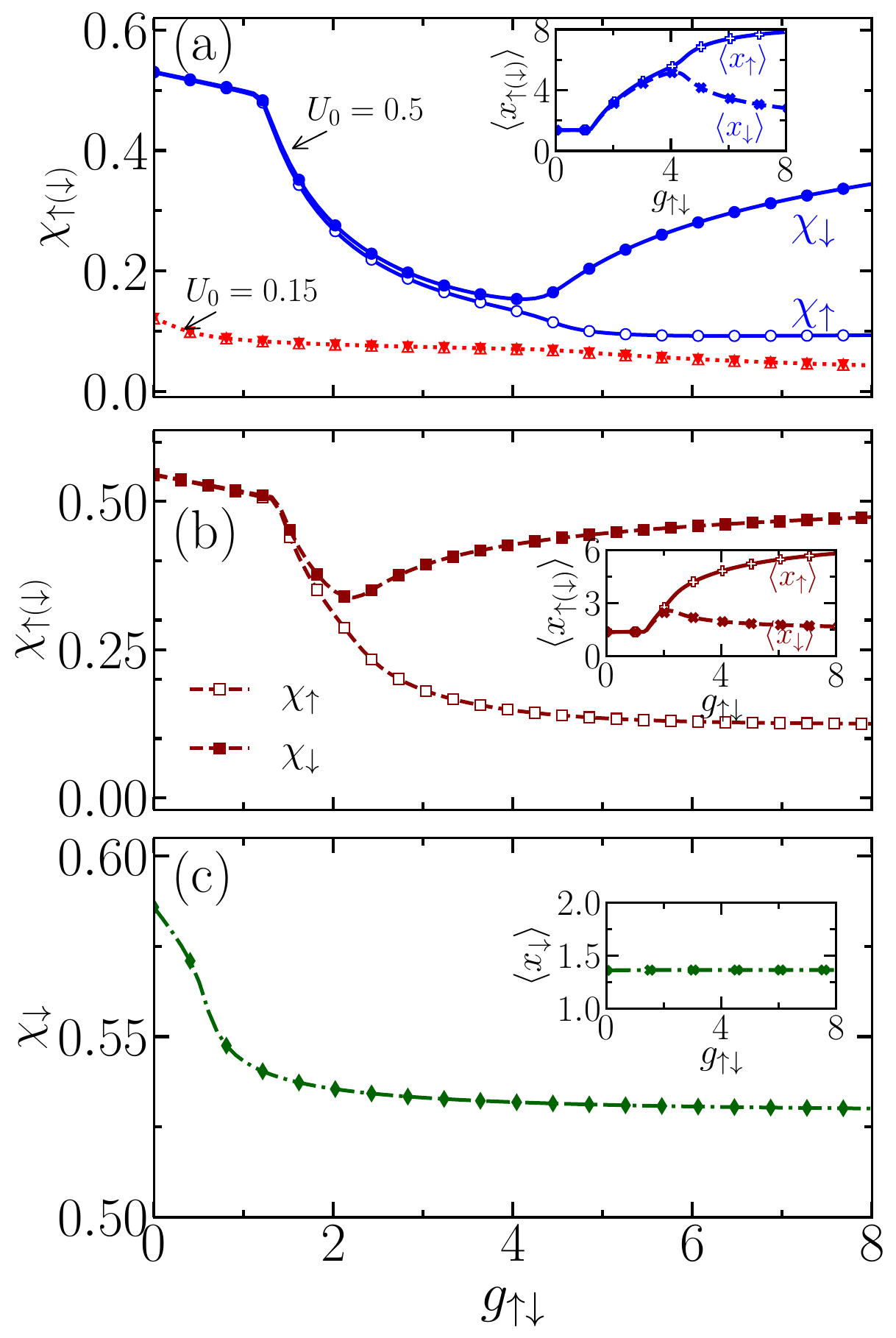}
 \caption{(a) Variation of the IPR $\chi_{\upar (\dar)}$ as a function of $g_{\upar\dar}$ at two different disorder strengths as : $U_{0} = 0.15$, and $U_{0} = 0.5$ at $k_{L} = 0$. IPR at (b) $k_{L} = 0.5, U_{0} = 0.5$, and (c) $k_{L} = 1.5, U_{0} = 0.5$. In figure (c), only $\chi_{\dar}$ has shown as the population in the spin-up component is zero. Here, the other parameters are $g_{\upar \upar} = 0.1, g_{\dar\dar} = 0.05$, and $\Omega = 0.5$. Inset figures show the center of mass of the spin-up and spin-down condensate which gets separated beyond the threshold value of $g_{\upar \dar}$. The results are shown for realization R1.}
 \label{fig:chi-g12}
\end{figure}
%%%%%%%%%%%%%%%%%%%%%%%%%%%%%%%%%%%%%%%%%%%%%%%%%%%%
%%%%%%%%%%%%%%%%%%%%%%%%%%%%%%%%%%%%%%%%%%%%%%%%%%%%
\begin{figure}[!htp]
\centering
\includegraphics[width=\linewidth]{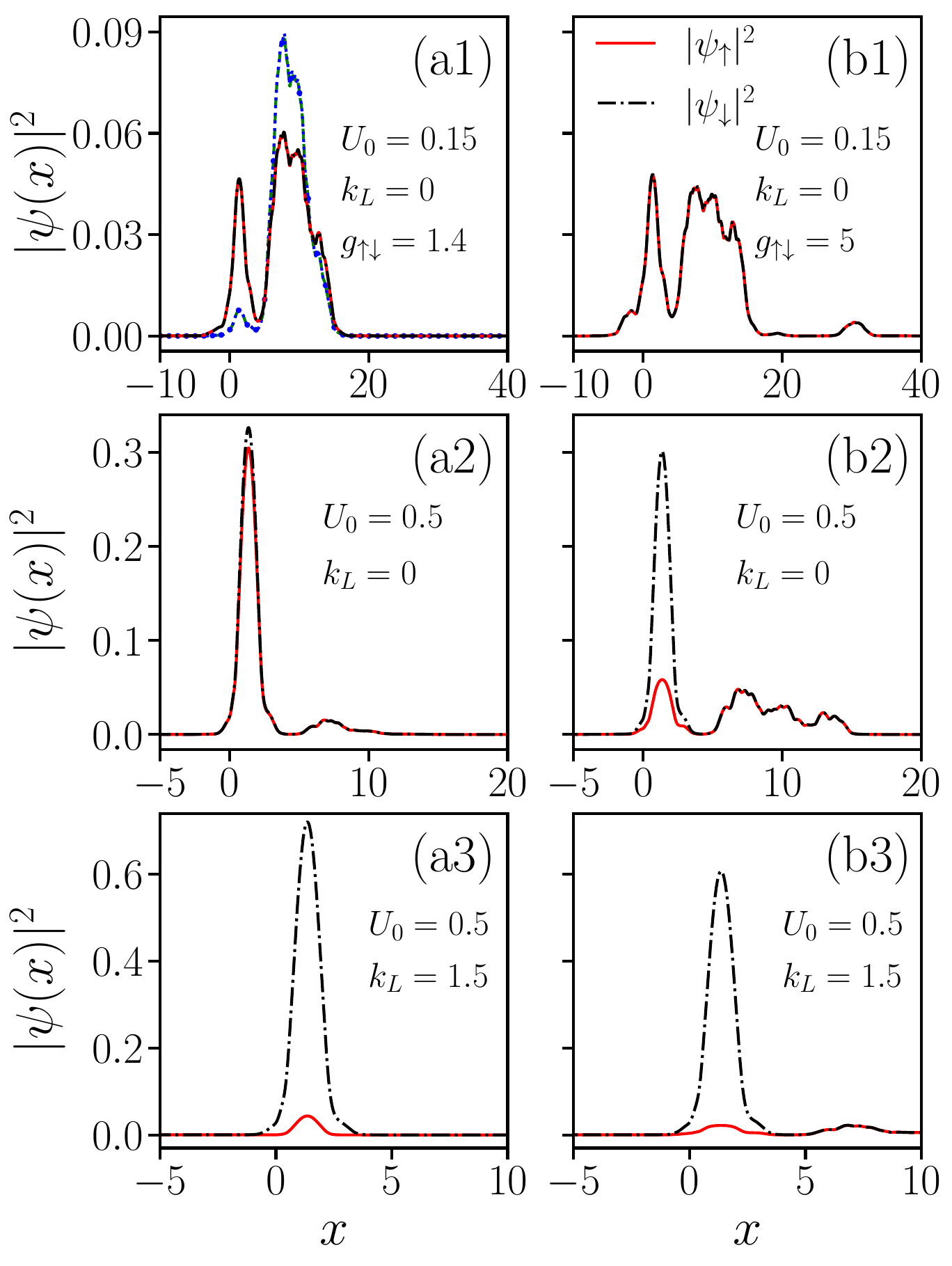}
\caption{Condensate density of up (solid red line), and down (dash-dotted black line) components for different disorder and SO coupling strengths as $U_{0} = 0.15, k_{L} = 0$ (a1, b1), $U_{0} = 0.5, k_{L} = 0$ (a2, b2), and $U_{0} = 0.5, k_{L} = 1.5$ (a3, b3) at inter-species interaction $g_{\upar\dar} = 1.4$ (left column), and $g_{\upar\dar} = 5.0$ (right column). The intra-species interactions are $g_{\upar \upar} = 0.1$, $g_{\dar\dar} = 0.05$, and $\Omega = 0.5$. In panel (a1) the density $\vert \psi_{\upar}\vert ^2 = \vert \psi_{\dar}\vert ^2$ (blue and green dashed lines) at $g_{\upar \dar} = 0$  is plotted for comparison.}
\label{fig:density-U0-0p15-0p5-g12}
\end{figure}
%%%%%%%%%%%%%%%%%%%%%%%%%%%%%%%%%%%%%%%%%%%%%%%%%
\begin{figure}
 \centering
 \includegraphics[width=\linewidth]{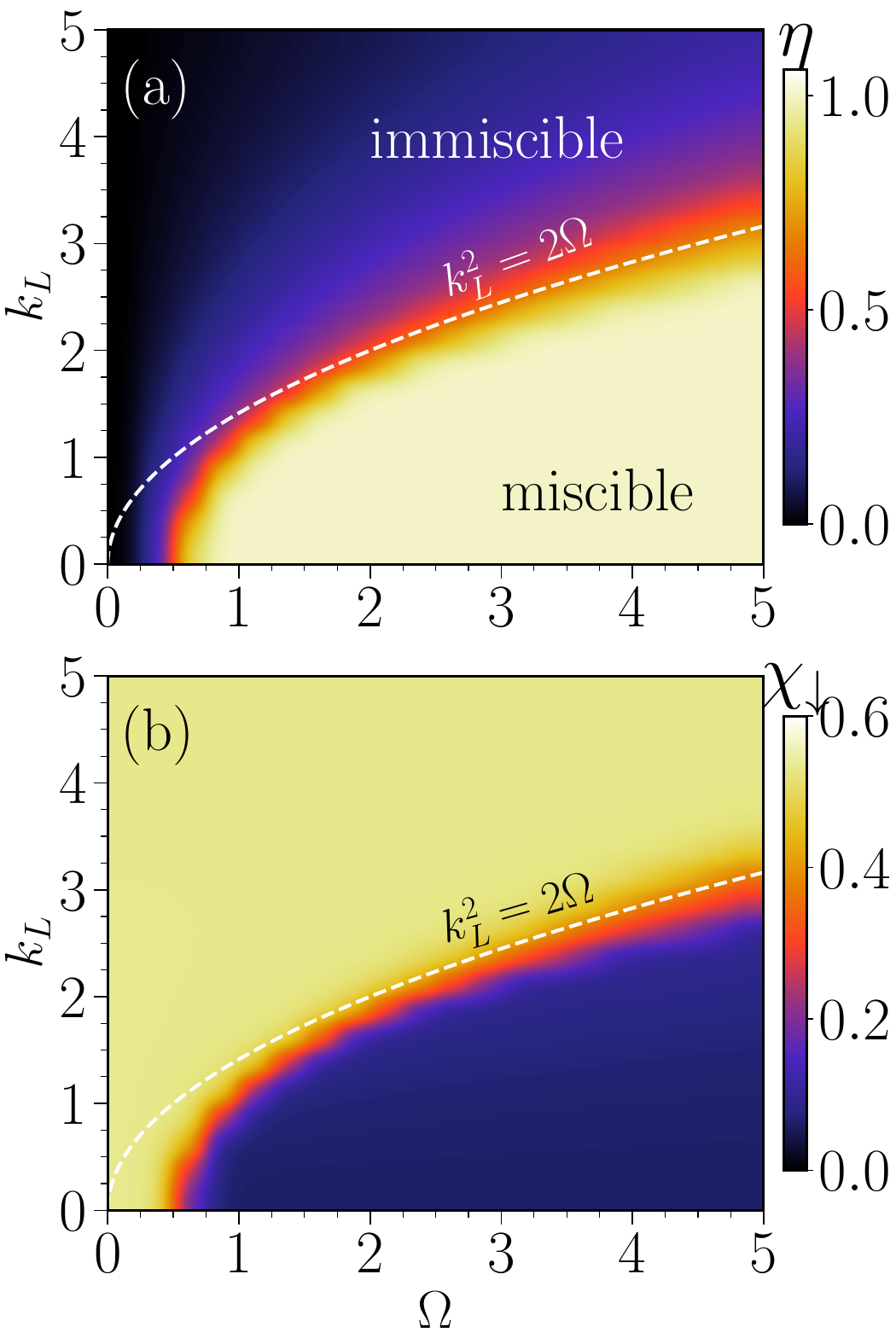}
 \caption{Pseudocolor representation of (a) spin miscibility $\eta$, and (c) IPR of spin-down $\chi_{\dar}$ in the ($\Omega,k_{L}$) plane at $g_{\upar \upar} = 0.1, g_{\dar\dar} = 0.05, g_{\upar\dar} = 5.0$. The condensate undergoes a transition from miscible ($\eta \approx 1$) to immiscible ($\eta \ll 1$) region where two regions are separated by following the $k_{L}^2 = 2\Omega$ relation as marked by the green line. While transitioning from miscible to immiscible region, the $\chi_{\dar}$ increases beyond $k_{L}^2 > 2\Omega$. Here, the disorder strength is $U_{0} = 0.5$ for R1. }
 \label{fig:phase-kL-Omega-g12-5p0}
\end{figure}
%%%%%%%%%%%%%%%%%%%%%%%%%%%%%%%%%%%%%%%%%%%%%%%%%%%

So far, our results revealed that the interplay of the disordered potential and the imbalance between intra-species interaction leads to the exchange of population from one component to another. Here we will show that nonlinearities can lead to stronger localization of one component at the expense of weaker localization of the other in the presence of a sufficiently strong disorder.

To understand the role of the inter-species interaction $g_{\upar\dar}$ we begin with the populations, followed by the variation of the IPR with $g_{\upar\dar}$ at the Rabi coupling $\Omega = 0.5$ for realization R1. Note that, the results for R2 exhibit similar qualitative features as those for R1. Therefore, in what follows we will only discuss the results pertaining to the R1 case. The populations $N_{\upar (\dar)}$ presented in Fig.~\ref{fig:Natoms-g12} show that for the broadly localized condensate at $U_{0} = 0.15, k_{L} = 0$, it remains the same across all the range of $g_{\upar\dar}$ where small $\chi$ leads to $g_{\upar\dar}\chi < 2\Omega$, rendering the components in the miscible state. In contrast, the narrow localized condensate at $U_{0} = 0.5$ shows that the populations in both components remained approximately constant at $N_{\upar} \approx N_{\dar} \approx 0.5$ only up to $g_{\upar\dar} \leq 4.6$ and then shows an increase in $N_{\dar}$ and corresponding decrease in population in spin-up component $N_{\upar}$ \sks{in case of $k_{L} = 0$. Similarly, the inclusion of $k_{L} = 0.5$ (brown dashed line), $1.5$ (green dash-dotted line) decrease the threshold value of interspecies interaction $g_{\upar \dar}$ for the separation of the condensate population. For $k_{L} =0.5,$ and $k_{L} = 1.5$ the populations in both component remain nearly constant at $0.5$ upto $g_{\upar \dar} \leq 2.1$, and $ g_{\upar \dar} \leq 0.5$, respectively.}  

Next, in Fig.~\ref{fig:chi-g12}(a), we show the variation of the IPR $\chi_{\upar (\dar)}$ with $g_{\upar\dar}$ at $U_{0} = 0.15$ and $U_{0} = 0.5$ by keeping the coupling parameters at $k_{L} = 0$ and $\Omega = 0.5$. For $U_{0} = 0.15$, one can notice the decreasing behavior of $\chi_{\upar} = \chi_{\dar}$ indicates the considerable broadening of the condensate. Conversely, for $U_{0} = 0.5$, initially, the IPR for both components remains the same and exhibits a decreasing trend until $g_{\upar\dar} \approx 4.6$. However, as $g_{\upar\dar}$ increases further, $\chi_{\dar}$ starts to increase due to the higher condensate density. The increase in $\chi_{\dar}$ is complemented by the decreasing behavior of $\chi_{\upar}$. This observation aligns well with the findings in Fig.~\ref{fig:Natoms-g12}. 

Notice that a considerable separation of $\chi_{\upar}$ and $\chi_{\dar}$ for 
 $U_{0} = 0.5$ starts at $g_{\upar\dar}\chi$ close to $2\Omega=1$, corresponding to the condition (i) formulated after Eq. \eqref{eq:Evarepsilon}. For the broadly localized state at $U_{0} = 0.15$, this product is always considerably less than 1, and the separation does not occur. The difference in the behavior of $\chi_{\upar}$ and $\chi_{\dar}$ is attributed to strongly different populations $N_{\upar}$ and $N_{\dar}$ with $N_{\upar}\ll\, N_{\dar}$ and their spatial separation since in the GPEs we have cross-interaction terms $g_{\upar\dar}\vert \psi_{\dar}\vert ^{2}\psi_{\upar}$ and $g_{\upar\dar}\vert \psi_{\upar}\vert ^{2}\psi_{\dar}$ in Eq.~(\ref{eqn1(a)}) and (\ref{eqn1(b)}), respectively. As a result, at a large $g_{\upar\dar}$, $\chi_{\upar}$ tends to almost constant while $\chi_{\dar}$ still decreases. 
 
 Figures~\ref{fig:Natoms-g12} and \ref{fig:chi-g12} show that the critical value of $g_{\upar\dar}$ causing spin polarization and the IPR difference decreases upon increasing the $k_{L}.$ This behavior corresponds to Eq.\eqref{eq:sigmax}, where at a strong SO coupling the role of the Rabi interaction rapidly decreases and the effect of cross-repulsion becomes dominant. Also, the $g_{\upar\dar}$ repulsion produces state with $\langle\sigma_{z}\rangle=N_{\upar}-N_{\dar}.$ At a sufficiently strong $g_{\upar\dar}$ one obtains $N_{\upar}-N_{\dar}\approx\,-1$ and purity $P\approx\,1.0$ in Fig.~\ref{fig:Natoms-g12}(c). For example, at $k_{L} = 0.5$ in Fig.~\ref{fig:chi-g12}(b), the $\chi_{\dar}$ starts increasing with respect to $\chi_{\upar}$ beyond $g_{\upar\dar} \approx 2.0,$ which is perfectly complemented by the populations in Fig.~\ref{fig:Natoms-g12}. Furthermore, the IPR at $k_{L} = 1.5$ [see Fig.~\ref{fig:chi-g12}(c)] exhibits similar trend. Since with the increase in $g_{\upar\dar}$ one reaches $N_{\upar}\ll\,1,$ we present only $\chi_{\dar}$ in this case.
 
 The role of $k_{L}$ can be formulated in terms of the joint effect of self-interaction and SO coupling. At a given choice of the intra-spin nonlinearities, they produce a state with $\langle\sigma_{z}\rangle<0$ resulting in $\langle\sigma_{z}\rangle\approx\,-1$ for a sufficiently large $g_{\dar\upar}.$ This cooperates with the effect of SO coupling in producing states with $\langle\sigma_{z}\rangle\approx\pm\,1$ with the corresponding energy contribution $-k_{L}^{2}/2.$ Thus, a strong SO coupling leads to the formation of the $\langle\sigma_{z}\rangle\approx\,-1$ states at a smaller $g_{\dar\upar},$ corresponding to the presented Figures.

In Fig.~\ref{fig:density-U0-0p15-0p5-g12}, we show the spin-projected condensate density at $g_{\upar\dar} = 1.4$ (left column), and $g_{\upar\dar} = 5.0$ (right column). At a weak disorder $U_{0} = 0.15$ (a1, b1), broadly localized condensates remain perfectly miscible ($\eta \approx 1$) for any value of interactions. In contrast to that, for the narrow localized state at $U_{0} = 0.5$, interaction leads to separate spin-up and spin-down components decreasing the spin miscibility. The induced localization of the spin-down component occurs by transferring the density mostly from the void at $\langle x\rangle\approx\,1.3$. Lastly, the inclusion of the $k_{L}$ enhance the transfer of density from one component to another into the void, as shown in the IPR in Fig.~\ref{fig:chi-g12}(b,c). 

 \sks{In general, the Figures \ref{fig:Natoms-g12}-\ref{fig:density-U0-0p15-0p5-g12} highlight  the qualitative picture for the ground state localization in a random 
potential.} The self-interaction energy due to the inter-spin repulsion can be minimized by diminishing the product $N_{\upar}N_{\dar}$ and the spatial overlap of $\psi_{\upar}$ and $\psi_{\dar}.$ This repulsion leads to the localization of different spin species in different regions of $V(x)$ such as voids and basins corresponding to the common chemical potential of these components in Eqs. (\ref{eqn1(a)}) and (\ref{eqn1(b)}) at the stationary state. Thus, increasing inter-species interaction induces localization of one component displaced with respect to the other.  With the increase in $g_{\dar\upar}$ at $g_{\upar\upar}>g_{\dar\dar}$ the BEC is accumulated in the spin-down state as $\chi_{\dar}$ and $\langle x_{\dar}\rangle$ gradually approach their values for the non self-interacting BEC. This picture works for initially sufficiently well-localized BEC in the void while the initially broadly localized state inside the basin with a small IPR remains inside the basin and broadens with the increasing inter-spin repulsion. Spin-orbit coupling enhances the effect of the $g_{\upar\upar}-g_{\dar\dar}$ difference and leads to the separation of spin-up and spin-down components at a weaker $g_{\upar\dar}$ interaction.
%%%%%%%%%%%%%%%%%%%%%%%%%%%%%%%%%%%%%%%%%%%%%%%%%%%%%%%%%%%%%%%
\begin{figure}[!ht]
 \centering
 \includegraphics[width=\linewidth]{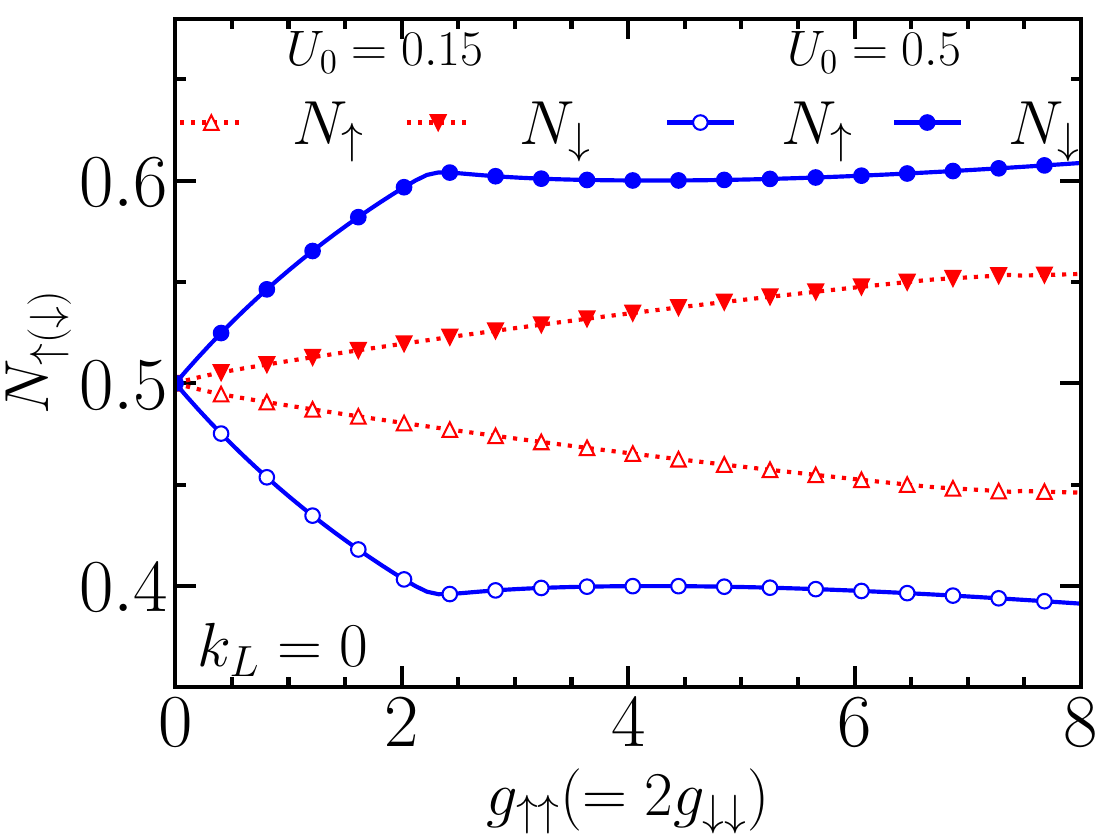}
 \caption{Populations of spin-up $N_{\upar}$ (open markers), and spin-down $N_{\dar}$ (filled markers) components as a function of intra-species interaction $g_{\upar \upar}(=2g_{\dar\dar})$ at $k_{L} = 0$. The other parameters are $g_{\upar\dar} = 0$, and $\Omega = 0.5$. }
 \label{fig:Natoms-g11-g22-g12-0}
\end{figure}
%%%%%%%%%%%%%%%%%%%%%%%%%%%%%%%%%%%%%%%%%%%%%%%%%%%%%%%%%%%%%%%%%%%%%%%
%%%%%%%%%%%%%%%%%%%%%%%%%%%%%%%%%%%%%%%%%%%%%%%%%%%%%%%%%%
\begin{figure}
 \centering
 \includegraphics[width=\linewidth]{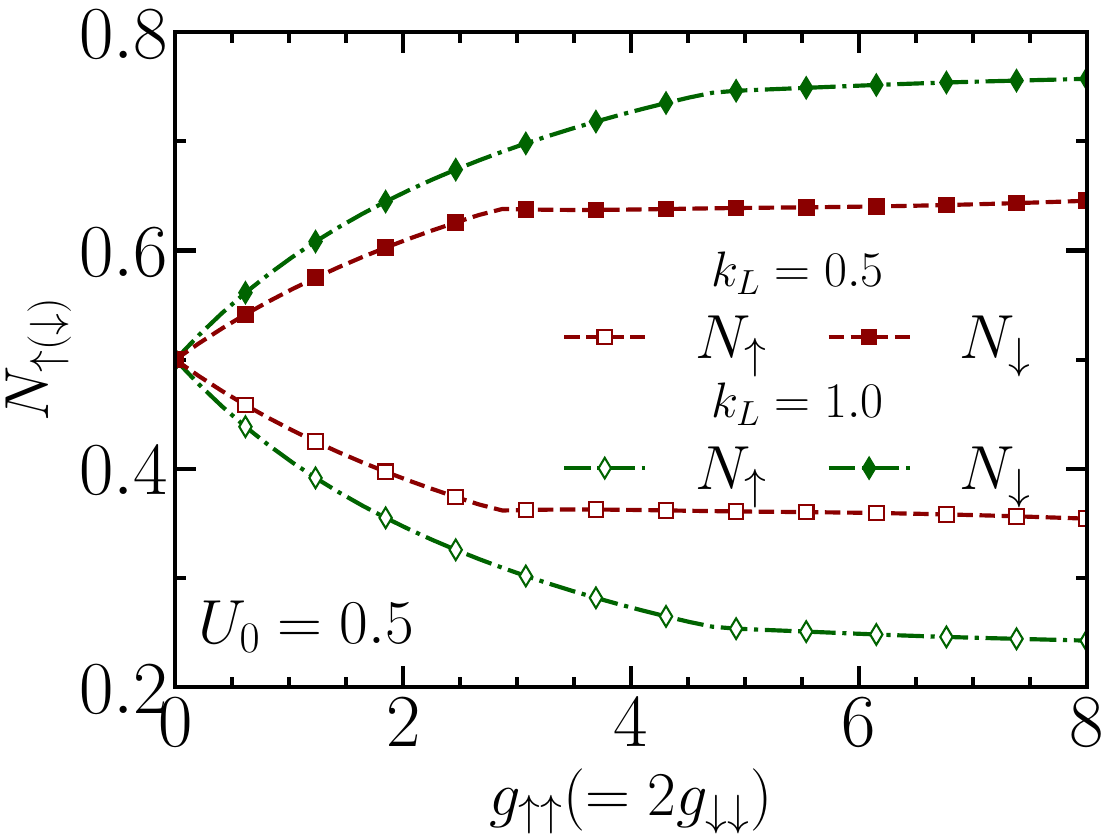}
 \caption{Populations of spin-up $N_{\upar}$ (open markers) and spin-down $N_{\dar}$ (filled markers) components as a function of intra-species interaction $g_{\upar \upar}(=2g_{\dar\dar})$ at $k_{L} = 0.5$, and $k_{L} = 1.0$. Here, the other parameters are $g_{\upar\dar} = 0$, and $\Omega = 0.5, U_{0} = 0.5$. Upon increasing $k_{L}$, leads to an increase in the rate of change of $N_{\upar (\dar)}$ with the variation of $g_{\upar \upar}$. }
 \label{fig:Natoms-g11-g22-g12-0-kL-0p5-1p0}
\end{figure}
%%%%%%%%%%%%%%%%%%%%%%%%%%%%%%%%%%%%%%%%%%%%%%%%%%%%%%%%

To further understand the effect of SO and Rabi coupling on induced localization, we calculate the miscibility and IPR for an extensive range of $k_{L}$ and $\Omega$ by keeping disorder strength $U_{0} = 0.5$ and $g_{\upar\dar} = 5.0$. In Fig.~\ref{fig:phase-kL-Omega-g12-5p0}, we present the miscibility and corresponding $\chi_{\upar(\dar)}$ in $(\Omega, k_{L})$ plane. The miscibility in panel (a) highlights the decrease in $\eta$ of localized condensate from the perfectly miscible state by following the $k_{L}^2 = 2\Omega$ relation for the non-zero value of $\Omega$. In the immiscible region ($k_{L}^2 > 2\Omega$), increasing $k_{L}$ predominantly causes the transfer of density from the spin-up to spin-down component, reducing the spin-up density to almost negligible value due to the condition $g_{\upar \upar} > g_{\dar \dar}$ [Fig.~\ref{fig:density-U0-0p15-0p5-g12}(a3,b3)]. As a result, $\chi_{\upar}$ in these immiscible regions becomes insignificant. However, the $\chi_{\upar}$, and $\chi_{\dar}$, shows that in the perfectly miscible region at $k_{L}^2 \ll 2\Omega$ the IPR remain $\chi_{\upar} \approx \chi_{\dar} \approx 0.08$, while in the immiscible region, the $\chi_{\dar}$ increases to $0.6$ results in localizing the down-component with respect to spin-up component as shown in Fig.~\ref{fig:phase-kL-Omega-g12-5p0}(b).

\subsection{Impact of unequal intra-species interaction with $g_{\upar\dar}=0$ on localization}

In the previous section, we explored the effect of inter-species interactions on the localized condensates by keeping the different small intra-species interactions constant. In this subsection, we investigate the effect of the intra-species interactions by keeping $g_{\upar\dar}=0$. 
 
\begin{figure}
 \centering
 \includegraphics[width=\linewidth]{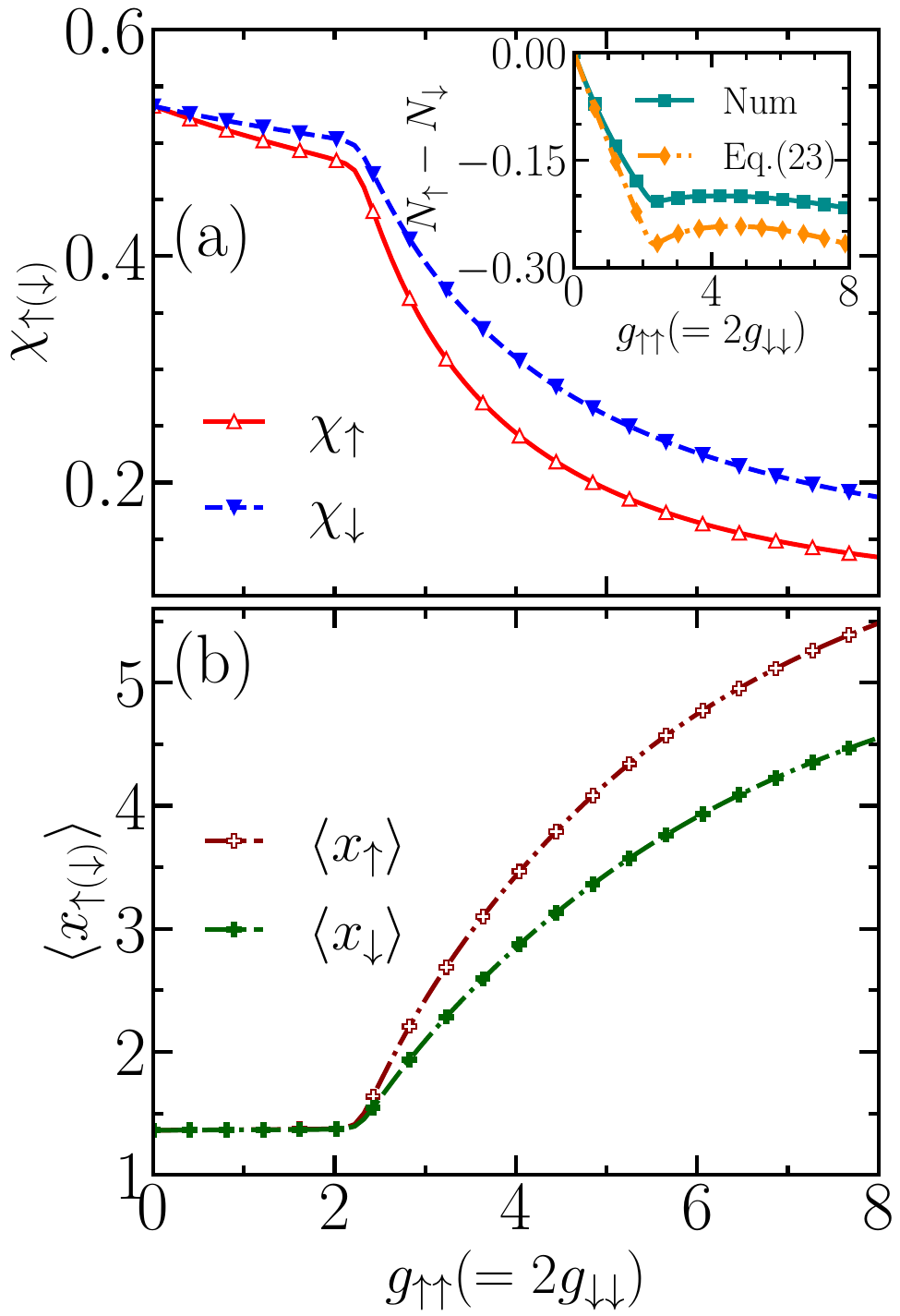}
 \caption{(a) Variation of the IPRs $\chi_{\upar (\dar)}$ as a function of the intra-species interaction $g_{\upar \upar} (=2g_{\dar\dar})$ for disorder strength $U_{0} = 0.5.$ With the increase in $g_{\upar \upar}$, the IPR in both components decreases, resembling the broadening of the condensate irrespective of disorder strength $U_{0}$. Other parameters are the same as Fig.~\ref{fig:Natoms-g11-g22-g12-0}. (b) The center of mass position for the up (red line) and down component (dashed blue line) as a function of the intra-species interaction $g_{\upar \upar}.$}
 \label{fig:chi-g11-g22-g12-0}
\end{figure}

 To investigate their effect, in Fig.~\ref{fig:Natoms-g11-g22-g12-0}, we plot the populations of spin-up and spin-down components as a function of $g_{\upar \upar}$ while the ratio between interactions kept as $g_{\dar\dar}/g_{\upar\upar} = 0.5$. One can notice that with the increasing $g_{\upar\upar}$, the populations in Fig.~\ref{fig:Natoms-g11-g22-g12-0} starts separating as $g_{\upar\upar} \neq 0$. The separation of $N_{\upar}$, and $N_{\dar}$ occurs for both narrow ($U_{0} = 0.5$) or a broadly localized state ($U_{0} = 0.15$). Notice that the splitting for $U_{0}=0.5$ is considerably stronger than that at $U_{0}=0.15$ since the value of $\chi$ at $U_{0}=0.5$ is considerably larger than that at $U_{0}=0.15$. At $k_{L}=0$, by using condition $\vert \varepsilon\vert \ll\,1$, that is $\chi(g_{\upar\upar}-g_{\dar\dar})\ll\,\Omega$, we obtain from Eq. \eqref{eq:Evarepsilon}:
\begin{align}\label{eq:vargg}
N_{\upar}-N_{\dar} = -\chi\frac{g_{\upar\upar}-g_{\dar\dar}}{4\Omega}, 
\end{align}
where at a weak self-interaction one can take $\chi$ for the $\phi_{0}$ function. This equation describes well the initial splitting in Fig.~\ref{fig:Natoms-g11-g22-g12-0}, valid up to $g_{\upar\upar}<2$, in agreement with Fig. \ref{fig:chi-g11-g22-g12-0} demonstrating $\chi_{\dar}$ and $\chi_{\upar}$ as a function of $g_{\upar\upar}$. 
%%%%%%%%%%%%%%%%%%%%%%%%%%%%%%%%%%%%%%%%%%%%%%%%%%%%%%%%%%%%%%%%%%%%
\begin{figure}
 \centering
 \includegraphics[width=\linewidth]{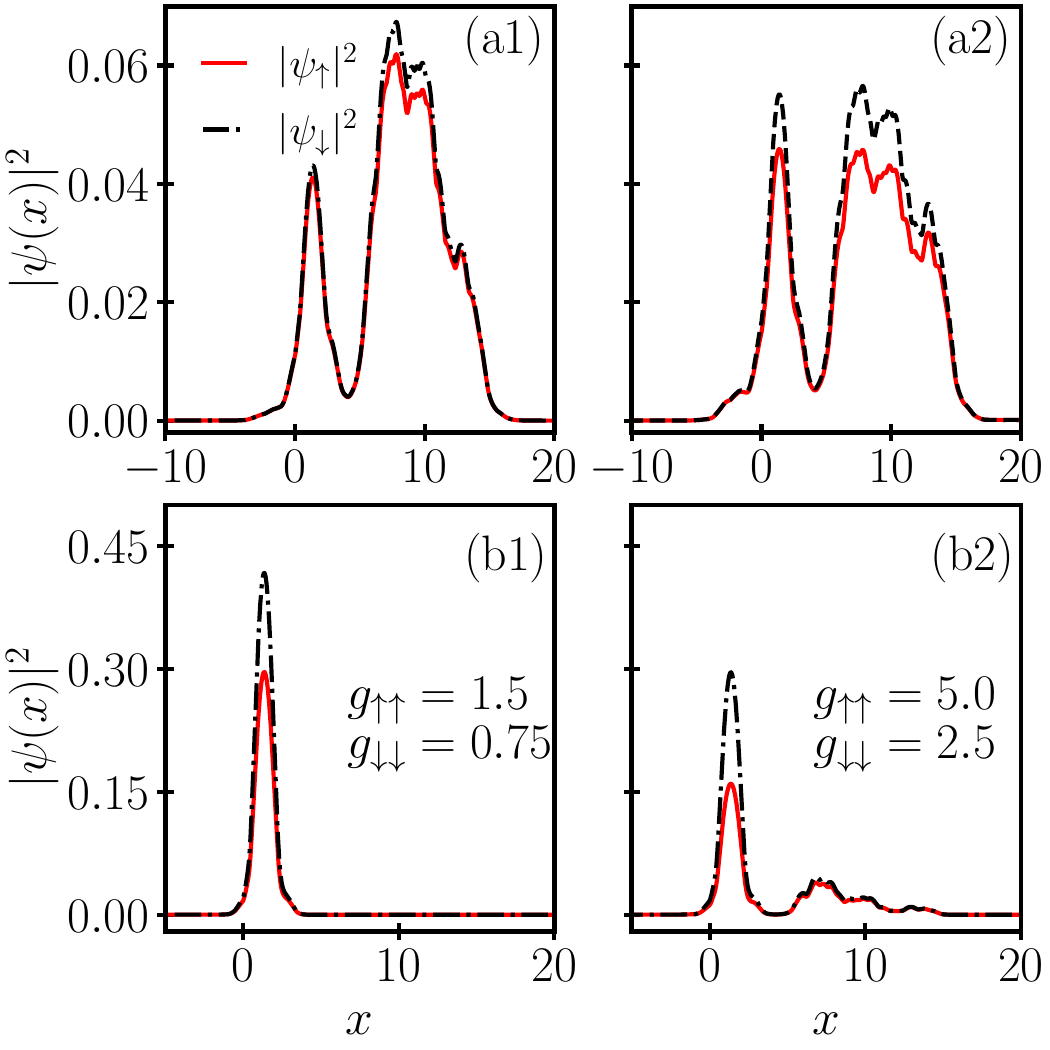}
 \caption{Ground state density profile of spin-up $\vert \psi_{\upar}\vert ^2$ (red solid line), and spin-down $\vert \psi_{\dar}\vert ^2$ (black dash-dotted line) for different inter-species interactions \sks{(a1, b1) $g_{\upar \upar} = 1.5, g_{\dar \dar} = 0.75$, and (a2, b2), $g_{\upar \upar} = 5.0, g_{\dar \dar} = 2.5$. }Here, the disorder strength is kept at $U_{0} = 0.15$ (in top panel), and $U_{0} = 0.5$ (in bottom panel). The other parameters are $g_{\upar\dar} = 0$, and $k_{L} = 0, \Omega = 0.5$ for realization R1.}
 \label{fig:den-sub-U0-0p15-0p5-g12-0}
\end{figure}
%%%%%%%%%%%%%%%%%%%%%%%%%%%%%%%%%%%%%%%%%%%%%%%%%%%%%%%
 In addition to that, the inclusion of $k_{L} = 0.5, 1.0$ [see Fig.~\ref{fig:Natoms-g11-g22-g12-0-kL-0p5-1p0}] enhance the separation of $N_{\upar (\dar)}$ between the components. For very high $g_{\upar \upar}$ values, the populations for both components saturate at a constant value. 
%\textcolor{red}{
%In Fig.~\ref{fig:Natoms-g11-g22-g12-0-kL-0p5-1p0} we show the population of the up and down spin component as a function of the intra-species interaction ($g_{\uparrow\uparrow}=2g_{\downarrow\downarrow}$) for $k_{L}=0.5$ and $k_{L} = 1.0$ considering the $\Omega=0.5$ and $U_{0}=0.5$. For both the $k_{L}$ the population for the down component ($N_{\uparrow}$) increases linearly on the expense of linear decrease of the up component population with the increase in the intra-species interaction. The population further gets saturated to a constant value as the interaction increases beyond a threshold value. The threshold frequency gets lowered upon decreasing the $k_{L}$. 
%}

Next, Fig.~\ref{fig:den-sub-U0-0p15-0p5-g12-0} illustrates the spatial variation of the condensate density for spin-up and spin-down components at different intra-species interaction as \sks{(a1, b1) $g_{\upar \upar} = 1.5, g_{\dar \dar} = 0.75$, and (a2, b2) $g_{\upar \upar} = 5.0, g_{\dar \dar} = 2.5$ with disorder strength $U_{0} = 0.15$ (upper panel), and $U_{0} = 0.5$ (lower panel)}. Here, the separation of two components occurs even when the condensate is broadly localized at the basin. As the intra-species interaction $g_{\upar \upar}$ increases, the miscibility decreases, and both components of the condensate start to spread toward the basin, eventually leading to delocalization as characterized also with the IPRs and centers of mass $\langle x_{\upar (\dar)}\rangle$ in Fig.~\ref{fig:chi-g11-g22-g12-0}(a) and \ref{fig:chi-g11-g22-g12-0}(b). 

A comparison of Figs.~\ref{fig:Natoms-g11-g22-g12-0-kL-0p5-1p0}-\ref{fig:chi-g11-g22-g12-0} with Figs.~\ref{fig:Natoms-g12}-\ref{fig:density-U0-0p15-0p5-g12} illustrates similarities and differences between the effects of intra- and inter-spin interactions. At nonzero $g_{\dar\dar}$ and $g_{\upar\upar}$ a sufficiently strong Rabi coupling forms a highly miscible state with $\eta\approx\,1$ localized either in a void or in a basin of the random potential. The choice of the localization position can be understood from Eq. \eqref{eq:Evarepsilon0} by choosing the minimum of $\tilde{\epsilon}_{p|\rm lin}+\chi_{p}\left(g_{\dar\dar}+g_{\upar\upar}\right)/8,$ where index $p$ corresponds to the void or basin, respectively. 
In a simplified analysis, we can assume that $\tilde{\epsilon}_{v|\rm lin}\approx \pi^{2}/(2l^{2}),$ $\tilde{\epsilon}_{b|\rm lin}\approx \langle V(x)\rangle$ with $\chi_{v}\approx 3/(2l)$ and $\chi_{b}\ll\,\chi_{v},$ where $l$ is the void width. Thus, the BEC is localized in the void if 
\begin{equation}\label{eq:voidbasin}
\frac{\pi^{2}}{2l^{2}}+\frac{3}{16l}\left(g_{\dar\dar}+g_{\upar\upar}\right)\lesssim\,\langle V(x)\rangle
\end{equation}
and in the basin otherwise, in agreement with our numerical results.
Although, at $g_{\upar\dar}=0,$ there is no direct interaction between the spin components, their separation in a random potential occurs since at $g_{\upar\upar}>g_{\dar\dar}$ given parameters of the system we consider, the spin-up BEC fraction energetically prefers to localize in the basins leaving $N_{\dar}$ in the voids to keep their common chemical potential. The degree of this separation is controlled by the Rabi coupling making it prohibited at sufficiently large $\Omega.$ Thus, the possible separation of the spin components here is caused not by their repulsion but by preferable localization dependent on the spin-dependent self-interaction.  
At a sufficiently large $g_{\upar\upar}-g_{\dar\dar}$ one can obtain $N_{\dar}\gg\,N_{\upar}.$ The self repulsion in each spin component decreases their IPRs $\chi_{\dar}$ and  $\chi_{\upar},$ and the overlap of $|\psi_{\dar}|$ and $|\psi_{\upar}|$ increases. The effect of SO coupling, enhancing the spin disproportion, is similar for both types of self-interaction.

%%%%%%%%%%%%%%%%%%%%%%%%%%%%%%%%%%%%%%%%%%%%%%%%%%%%%%%%%
%%%%%%%%%%%%%%%%%%%%%%%%%%%%%%%%%%%%%%%%%%%%%%%%%%%%%
%%%%%%%%%%%%%%%%%%%%%%%%%%%%%%%%%%%%%%%%%%%%%%%%%%%%%
\begin{figure}
\centering
\includegraphics[width=\linewidth]{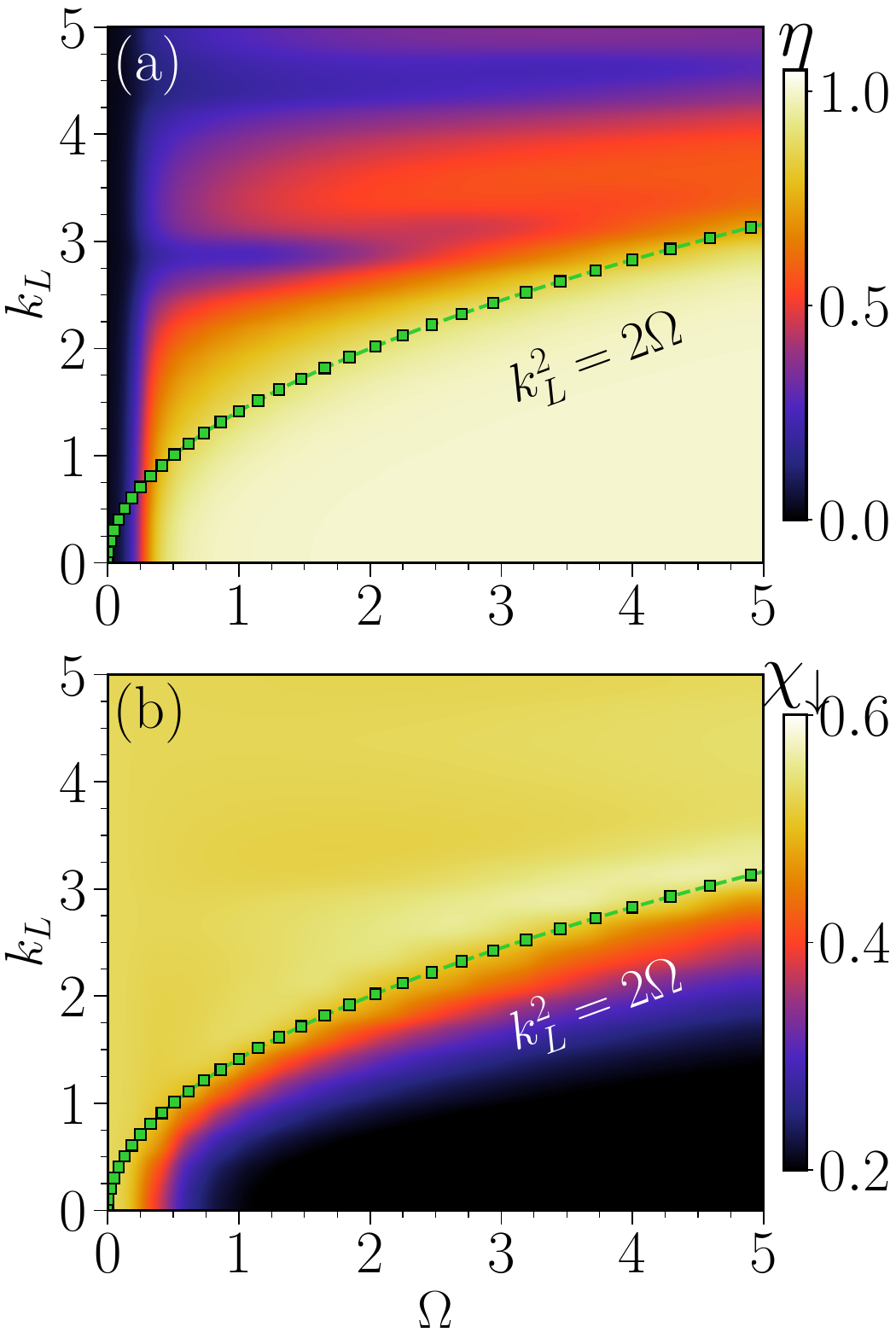}
\caption{Pseudocolor representation of (a) spin-miscibility $\eta$, (b) IPR of spin-down $\chi_{\dar}$ components in ($k_{L}, \Omega$) plane at intra-species interactions $g_{\upar \upar} = 5.0, g_{\dar\dar} = 0.5g_{\upar \upar}$, and inter-species interaction $g_{\upar\dar} = 0.0$. Here, the $k_{L}^2 = 2\Omega$ line is drawn to show the transition from broadly localized condensate in the basin to narrowly localized condensate towards the void. Here, the IPR for down component follows $k_{L}^2 = 2\Omega$ relation.}
\label{fig:phase-kL-Omega-g12-0-g11-5p0}
\end{figure}
%%%%%%%%%%%%%%%%%%%%%%%%%%%%%%%%%%%%%%%%%%%%%%%%%%%%%
%%%%%%%%%%%%%%%%%%%%%%%%%%%%%%%%%%%%%%%%%%%%%%%%%%%%%
\begin{figure}
 \centering
\includegraphics[width=\linewidth]{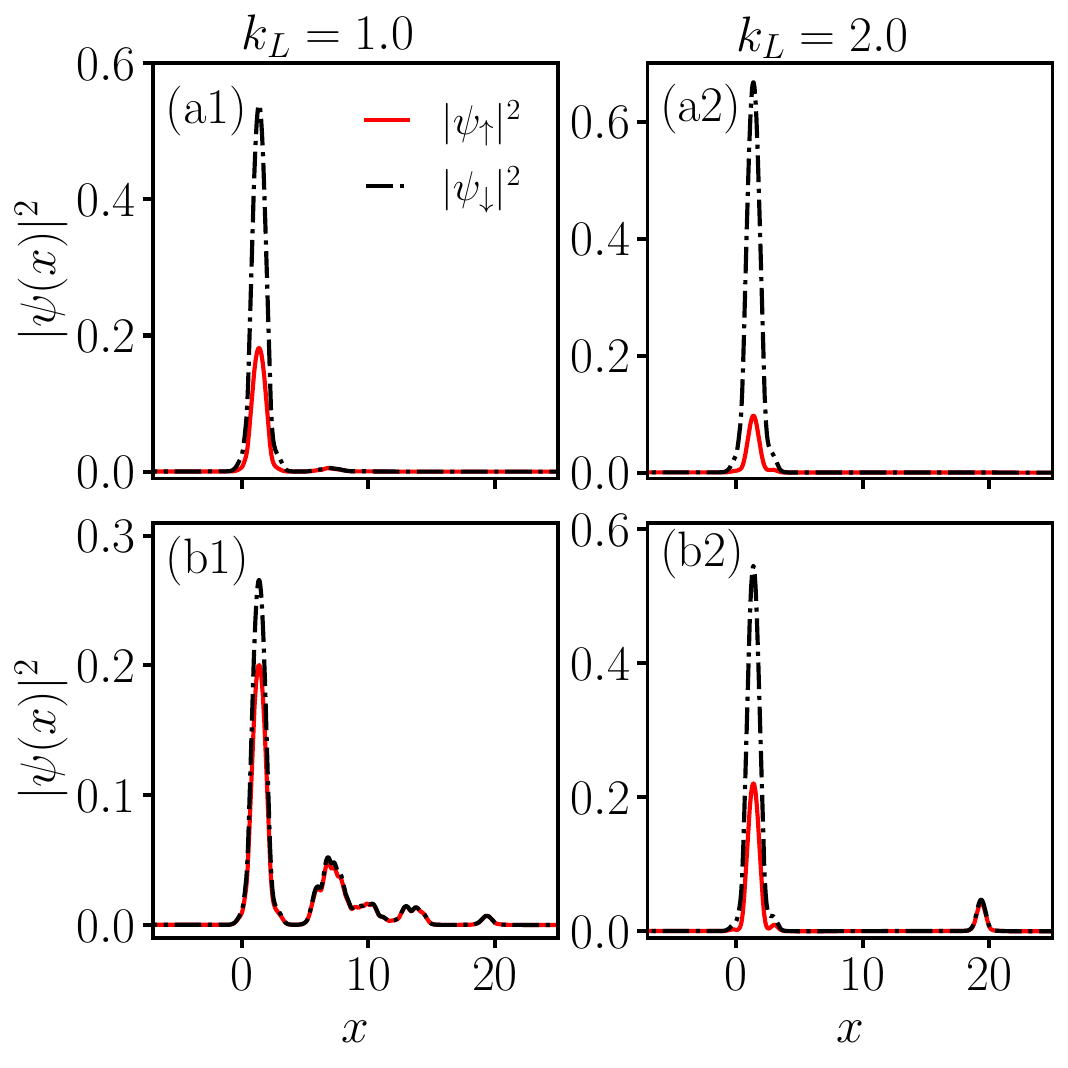}
 \caption{Ground state density profile of spin-up $\vert \psi_{\upar}\vert ^2$ (red solid line), and spin-down $\vert \psi_{\dar}\vert ^2$ (black dashed line) for different inter-species interactions (a1, b1) $k_{L} = 1.0$, and (a2, b2), $k_{L} = 2.0$. Here, the disorder strength is kept at $\Omega = 0.5$ (in the top panel), and $\Omega = 2.0$ (in the bottom panel). The other parameters are $g_{\upar\dar} = 0$, and $g_{\upar \upar} = 5.0$, $g_{\dar \dar} = 2.5$ and the disorder strength $U_{0} = 0.5$ for realization R1. }
 \label{fig:density-var-kL-g11-5}
\end{figure}
%%%%%%%%%%%%%%%%%%%%%%%%%%%%%%%%%%%%%%%%%%%%%%%%%%%%%%%%
% \begin{figure}[!ht]
%  \centering
%  \includegraphics[width=\linewidth]{density-kL-3p5-Omega-3p0.pdf}
%  \caption{Ground state density profile of spin-up $\vert \psi_{\upar}\vert ^2$ (red solid line), and spin-down $\vert \psi_{\dar}\vert ^2$ (black dashed line) for $k_{L} = 3.5, \Omega = 3.0$. The other parameters are $g_{\upar \upar} = 5.0, g_{\dar \dar} = 2.5, g_{\upar \dar} = 0$ and disorder strength $U_{0} = 0.5$ for realization R1}
%  \label{fig:density-kL-3p5-Omega-3p0}
% \end{figure}
%%%%%%%%%%%%%%%%%%%%%%%%%%%%%%%%%%%%%%%%%%%%%%%%%%%%%%%%
In Fig.~\ref{fig:phase-kL-Omega-g12-0-g11-5p0} we plot the miscibility and IPR in $(\Omega,k_{L})$ plane by keeping $g_{\upar \upar} = 5.0, g_{\dar\dar} = 2.5,$ and Rabi coupling $\Omega = 0.5.$
At a small $k_{L}$ and a relatively small $\Omega$ one obtains disproportion of $N_{\upar}$ and $N_{\dar}$ with $\eta\ll 1.$ With the increase in $\Omega,$ the role of the self-interaction with respect to the Rabi coupling decreases. Consequently, $\eta$ approaches 1, as shown in panel (a). The increase in the self-repulsion decreases both $\chi_{\dar}$ and $\chi_{\upar},$ as presented in the (b) panel.

With the increase in $k_{L},$ the panel (a) shows that the transition from miscibility to immiscibility requires $k_{L}\gtrsim\sqrt{2\Omega}.$ This is due to the involvement of spin-projected probability density distributions dependent on the disorder and self-interactions. Here the same spin repulsion decreases the IPRs of both spin components and increases their overlap and the spin miscibility. This, to increase the separation and decrease the miscibility, one needs a stronger SO coupling with $k_{L}^2$ exceeding $2\Omega$ by a value of the order of $g_{\dar\dar}\chi_{v}.$ However, the modification of $\chi_{\dar}$ roughly follows this boundary where $|\psi_{\dar}|^2$ shows transition from delocalized to localized spin-down BEC component in $k_{L}^2 \gtrsim 2\Omega$ region. Since at these system parameters, the condensate population for a spin-up component in the immiscible region is very low, we present only the results of $\chi_{\dar}.$ 

For a better understanding of the effect of SO coupling, we present in Fig.~\ref{fig:density-var-kL-g11-5} the spin-projected probability density distributions. This Figure shows that at a large $k_{L}$ the BEC is almost fully polarized in the spin-down state with $N_{\dar}\approx\,1$ localized in the void of the random potential. The reason for localization in the void rather than in the basin can be seen from the argument similar to Eq. \eqref{eq:voidbasin} since $\pi^{2}/(2l^{2})+3g_{\dar\dar}/(4l)\lesssim\,\langle V(x)\rangle.$

% \pagebreak
\section{Conclusions}\label{conclusion}

We have investigated the effects of the interplay between disorder, self-interactions, Rabi, and spin-orbit coupling on the localization of a Bose-Einstein condensate located in a one-dimensional random potential.  This potential produces two distinct types of the ground state localization: one in the void-like and the other one in the basin-like regions of the random potential. Both localization patterns are different from conventional Anderson localization attributed to the interference of matter waves scattered by different defects.  

Our study considered both non-self-interacting condensates and condensates with two types of nonlinear interactions, which lift Manakov's symmetry. Initially, we examined how SO coupling affects the localization of a linear condensate without self-interactions. In the absence of these interactions SO coupling significantly modifies the shape and spatial scale of the localized condensate, influencing its spatial positioning. An interesting manifestation of this effect is that SO coupling can move localized states from the basin to the void regions. 

By considering different types of self-interaction we observed that they have markedly distinct impacts on condensate behavior. These nonlinearities lead to pronounced spin-dependent effects related to the lifted Manakov's symmetry, resulting in the spatial redistribution of spin population and separation of spin components between the voids and basins of the random potential. The combined effects of SO coupling and self-interactions amplify the impact of nonlinearities, produce highly spin-polarized condensate, and lead to the localization of one spin component of the condensate at the expense of the other. In contrast to the formation of the stripe phase at a sufficiently SO coupling in the absence of disorder, here it causes localization of the condensate. This behavior has been identified by analyzing the probability density distributions, spin expectation values, spin miscibility, and purity.

\sks{The proposed model for the localization of quasi-1D spin-orbit coupled Bose-Einstein condensates in a random potential can be experimentally realized utilizing two hyperfine states of $^{87}$Rb atom considering pseudo spin-up atomic state as $\ket{\upar} \equiv \ket{F = 1, m_F = 0},$ and pseudo spin-down state $\ket{\dar} \equiv \ket{F = 1, m_F = -1}$ coupled using two Raman lasers with wavelength $\lambda = 804.1$ nm where the spin-orbit coupling $k_{L}$ can be tuned by varying the angle between the laser beams~\cite{Spielman:2011}.
The condensate is trapped under the strong transverse frequency $\omega_{\perp} \sim 10^{3}$ s$^{-1}$ with the characteristic length scale $a_{\perp}\sim 1$ $\mu$m and the characteristic energy scale 
 $\hbar\omega_{\perp}\sim 5$ nK. Variation of the angle between the beams permits to produce dimensionless  spin-orbit coupling $k_{L}$ and Rabi frequency $\Omega$ considered in the work in the interval from $\sim 0.1$ to $\sim 5.$ 
%and However to realize  dimensionless  in the range  $[0.1, 4]$ 
%the Raman laser intensity can be tuned in the range of $\tilde{\Omega}=2\pi\times [0.1-0.3]kHz$.
%the corresponding variations in the $\lambda_L$ can be made in the range between $73.91\mu$m and $1.84\mu$m. 

%For $N \sim 2 \times 10^4$ atoms  
 
In the absence of external magnetic field the interatomic scattering lengths $a_{\uparrow\uparrow},a_{\downarrow\downarrow},$ and $a_{\downarrow\uparrow},a_{\uparrow\downarrow}$ are of the order of $100a_{0}$ where $a_{0}$ is the Bohr's radius, and can be tuned by appropriate experimental setup \cite{Cornish:2000,Marte:2002,Erhard:2004,Egorov:2013}. For these interatomic scattering lengths and above given $a_{\perp},$ the resulting dimensionless interaction parameters $g_{\uparrow\uparrow},g_{\downarrow\downarrow},$ and $g_{\uparrow\downarrow},g_{\downarrow\uparrow}$ in the range used in our simulations can be achieved for the total number of atoms $\mathcal{N} <10^{4}.$ For larger $\mathcal{N},$ similar nonlinear interactions can be achieved at properly tuned smaller interatomic scattering lengths.  In laboratory experiment, the random potential is produced by using scattered coherent Gaussian laser light from the rough surface of a glass diffuser. The resulting intensity pattern is due to constructive and destructive interference of light from different scattering regions, forming a high contrast randomized distribution of the light intensity. The necessary condition for a good random potential is that the produced profile  must significantly extend the laser wavelength~\cite{clement:2006,Pasek:2017}. With an optimal number of scatterers relative to the diffuser size, destructive interference can generate the ``void'' regions, while alternating interference patterns may generate the ``basins''. The resulting dimensionless potential parameter $U_{0}$ can be tuned in the range from $\sim 0.05$ to $\sim 0.5$ ~\cite{clement:2006,Billy:2008}}.

Our work demonstrates that a broad variety of quantum states can be produced by the interplay of spin-dependent interactions and nonlinearities in a condensate located in a random potential. One of the future research directions may be an exploration of similar types of spin-dependent localization in higher spatial dimensions. It would be also interesting to extend the present study for more complex spinor systems such as spin-1 and spin-2 condensates. 
%\section{acknowledgments}
\begin{acknowledgments}
The work of E.Y.S. is supported through Grants No. PGC2018-101355-B-I00 and No. PID2021-126273NB-I00 funded by MCIN/AEI/10.13039/501100011033 and by the ERDF “A way of making Europe”, and by the Basque Government through Grant No. IT1470-22. We also gratefully acknowledge our supercomputing facilities Param-Ishan and Param-Kamrupa (IITG), where all the numerical simulations were performed. 
The work of PM is supported by MoE RUSA 2.0 (Physical Sciences -- Bharathidasan University). 
\end{acknowledgments}

\bibliography{references_random.bib}
\end{document}